\documentclass[useAMS, usenatbib, usegraphicx, twocolumn]{mn2e}

\input epsf
\usepackage{graphicx}
\usepackage{color}
\usepackage{amsfonts}
\usepackage{hyperref}

\newcommand{\adsurl}[1]{\href{#1}{ADS}}

\newcommand{\beq}{\begin{equation}}
\newcommand{\eeq}{\end{equation}}
\newcommand{\barr}{\begin{eqnarray}}
\newcommand{\earr}{\end{eqnarray}}

\newcommand{\new}[1]{#1}

\newcommand{\lesssim}{\mathrel{\hbox{\rlap{\lower.55ex\hbox{$\sim$}} \kern-.3em \raise.4ex \hbox{$<$}}}}
\newcommand{\gtrsim}{\mathrel{\hbox{\rlap{\lower.55ex\hbox{$\sim$}} \kern-.3em \raise.4ex \hbox{$>$}}}}
\title[Rotational spectroscopy of PAHs]{Rotational spectroscopy of interstellar PAHs}
\author[Y. Ali-Ha\"{\i}moud]{
       Yacine Ali-Ha\"{\i}moud\thanks{E-mail: yacine@ias.edu} \vspace{1.5mm}\\
Institute for Advanced Study, Einstein Drive, Princeton, New Jersey 08540}

\begin{document}

\date{\today}

\pagerange{\pageref{firstpage}--\pageref{lastpage}} \pubyear{2012}
\pagenumbering{arabic}
\label{firstpage}

\maketitle

\begin{abstract}

\new{Polycyclic aromatic hydrocarbons (PAHs) are believed to be ubiquitous in the interstellar medium. Yet, to date no specific PAH molecule has been identified. In this paper, a new observational
avenue is suggested to detect individual PAHs, using their rotational line emission at radio frequencies. Previous PAH searches based on
rotational spectroscopy have only targeted the bowl-shaped corannulene molecule, with the underlying assumption that other polar PAHs are triaxial and have a complex and diluted spectrum unusable for identification purposes. In this paper the rotational
spectrum of quasi-symmetric PAHs is computed analytically. It is shown that the asymmetry of planar, nitrogen-substituted symmetric PAHs is
small enough that their rotational spectrum, when observed with a resolution of about a MHz, has the appearance of a ``comb'' of evenly spaced stacks of lines. The simple pattern of these ``comb'' spectra allows for the use of matched-filtering
techniques, which can result in a significantly enhanced signal-to-noise ratio. Detection forecasts are discussed for regions
harbouring ``anomalous microwave emission'', believed to originate from the collective
PAH rotational emission. A systematic search for PAH lines in various
environments is advocated. If detected, PAH ``combs'' would allow to the conclusive and unambiguous identification of specific,
free-floating interstellar PAHs.}

\end{abstract}


\section{Introduction}

Polycyclic aromatic hydrocarbons (PAHs) are believed to be ubiquitous
in the interstellar medium (ISM). They are estimated to contain several percent of the total
interstellar carbon, which translates to a typical abundance of a few times $10^{-7}$ PAH per hydrogen atom for a characteristic number of $50$ carbon
atoms per molecule, making PAHs, as a family, one of the most abundant
molecular species in the ISM. They play important roles in the physics and
chemistry of the ISM, influencing, and sometimes controlling, the ionisation balance and the photoelectric heating (for a recent review, see
\citealt{Tielens_2008}; see also \citealt{Draine_book}). The presence of these large carbonaceous compounds in the ISM is strongly hinted at, if
not required, by various otherwise ``unidentified'' or ``anomalous''
emission and absorption features throughout the electromagnetic
spectrum. First and foremost, there is now general agreement for
attributing the infrared emission features at 3-19 $\mu$m to PAH vibrational
transitions following transient heating to high temperatures by
ambient ultraviolet (UV) photons \citep{Puget_1989, Allamandola_1989},
even though this is still the subject of occasional heated debates
\citep{Kwok_2011, Li_2012, Kwok_2013}. Second, electronic transitions of ionised
PAHs could be at the origin of the diffuse interstellar bands (DIBs) in
the visible \citep{Leger_1985, Crawford_1985, vdz_1985}, though the case is not
fully settled \citep{Tielens_2008}. The ``extended red emission''
could be the counterpart of the DIBs in emission, and PAHs have also been
suggested as its carriers \citep{Witt_2006}. Third, the best candidate for the cm-wavelength
``anomalous'' dust-correlated emission is rotational emission from
very small spinning dust grains \citep{DL98_short, DL98_long}, at the very least
for the lack of better candidates (with the exception of magnetic
dipole radiation suggested by \cite{DL_1999}, though a recent re-investigation of
the problem by \cite{Draine_2013} suggests typical radiation frequencies higher than the
observed anomalous emission), and also because
observations seem to fit well theoretical models, see for example
\cite{Scaife_2013} and references therein. These various aspects make a strong case for the PAH hypothesis,
further reinforced by their expected stability under the harsh
interstellar UV radiation if they contain $\sim$20 carbon atoms or more
\citep{Leger_1985}. And yet, specific PAHs have eluded
detection thus far, which is one of the main arguments of the
PAH-hypothesis opponents \citep{Kwok_2011, Kwok_2013}. It is indeed
difficult to extract information about individual carriers from the
infrared emission features, which are mostly due to nearest-neighbour
vibrations, and there is currently little data at the far-infrared
frequencies, which correspond to the bending modes of the skeleton and could trace
specific molecules \citep{Tielens_2008, Zhang_2010}.

Rotational spectroscopy is routinely used to identify molecules in
space, and it is natural to ask whether this technique can
be used for PAHs. Suggestions to do so can be found in a few places in the literature
\citep{Hudgins_2005, Tielens_2008, Hammonds_2011}, but so far the only PAH
that has been searched for using its rotational emission is the
polar, bowl-shaped corannulene molecule C$_{20}$H$_{10}$. Its
rotational constants were measured by \cite{Lovas_2005}, whose results
were used by \cite{Thaddeus_2006} and \cite{Pilleri_2009} to set an
upper bound to the fraction of interstellar carbon locked in
corannulene of $\sim 10^{-5}$ and $\sim 2 \times 10^{-6}$,
respectively, which are much smaller values than the estimated $\sim 10\%$
fraction of carbon locked in PAHs. These results only apply to
specific lines of sight, however, and to a very specific PAH which may
simply not be efficiently formed in the ISM or may be too small to be
stable under the harsh UV background. A search for rotational
transitions of other PAHs has never been carried out, however, due to the following \emph{a
  priori} difficulties. Firstly, the number of possible
configurations with a few tens of carbon atoms is a priori
gigantic \citep{Salama_1996}, and it could very well be that no
specific species is abundant enough that it could ever be detected,
with any method (this could explain the very low upper bounds on the
abundance of corannulene). That being said, the more compact, ``pericondensed''
configurations are more stable than the more
linear, ``catacondensed'' forms \citep{Ricca_2012}, and it is conceivable that only
a limited number of stable species dominate the PAH population (``grandPAHs'', in
the words of A.~Tielens). In fact, comparison of theoretical infrared
spectra with observations do suggest that interstellar PAHs are rather
compact and symmetric \citep{Bausch_2009}. Moreover,
the presence of a limited number of features, both in the mid-IR
spectrum and in the DIBs, suggests that their population is, indeed
(provided they are the carriers of these features), dominated by a
relatively small number of definite species \citep{Salama_1996,
  Boersma_2010, Ricca_2012}, though this statement has never been made fully
quantitative. Secondly, ideal PAHs are not necessarily polar, especially
the highly symmetric ones that are the most likely to be
overabundant. If this is the case they cannot radiate through
rotational transitions. Interstellar PAHs are however likely to have
some form of impurity (dehydrogenation, super-hydrogenation,
substitution...) that would endow them with a permanent dipole moment. Finally, the PAHs that are polar are in general triaxial
molecules (except for very special cases like that of corannulene), either because of their
intrinsic geometry, or because of the very impurities that make them
polar and break their symmetry even if they are initially perfectly
symmetric. The power radiated by large triaxial molecules is in general diluted between a very large number of weak lines, which may
disappear in the grass and be of little use for identification
purposes \citep{Tielens_2008}. This last point, however, has never
been made quantitative, which is one of the goals of the present work.

In this paper, we first compute the emission spectrum of
\emph{quasi-symmetric} rotors using quantum mechanical perturbation theory (Section \ref{sec:quasi-sym}). We show that for a small enough degree
of asymmetry, the rotational spectrum of a planar PAH, observed with a
few MHz resolution, has the appearance of a ``comb'' with
evenly spaced ``teeth'', each one of which being really a stack of a large number of
transitions that fall at nearly identical frequencies. We then
quantify the level of asymmetry of realistic PAHs with various
imperfections in Section \ref{sec:real-PAH}. We show in particular
that nitrogen-substituted PAHs (PANHs), which are strongly polar and are believed to be a significant
fraction of the interstellar PAH population \citep{Hudgins_2005}, do
remain symmetric enough to have a ``comb'' spectrum. They therefore
make a promising target for PAH rotational spectroscopy. The very
distinctive pattern of such ``comb'' spectra allows for the use of
matched-filtering techniques to blindly search for specific PAHs with
unknown size, as we describe in Section \ref{sec:comb-fitting}. Such techniques could
allow for the statistical detection of combs, even if none of the lines are strong enough to be detected on their own. We make rough detection forecasts in Section \ref{sec:forecast}, based
on the observed amplitude of the ``anomalous microwave emission''
(AME), assuming it is the broadband and collective rotational emission from PAHs. Our estimates indicate that the detection of
individual PAHs with current radio telescopes may be challenging but
remains possible provided the fraction of PAHs in a particular species
is large enough (typically of order a percent, but possibly less). We advocate for a systematic search of PAH
rotational lines in large bandwidth, few MHz resolution radio spectra of regions with known
anomalous microwave emission, as well as in regions with known or potential large PAH
abundances. Protoplanetary or planetary nebulae also make interesting
targets, and the detection of individual PAHs in these regions would shed light on the route to large organic
molecules and constrain chemical models \citep{Kwok_2004}. Detection
of PAH rotational lines would be a smoking-gun confirmation of their
existence as free-floating molecules in the interstellar medium, and would allow to finally
identify individual species, a nagging missing piece to the otherwise
very appealing PAH hypothesis.

\section{Rotational line emission from quasi-symmetric planar PAHs} \label{sec:quasi-sym}

\subsection{Motivations}

An ideal planar rigid symmetric top emits a series of evenly spaced
rotational lines, with a line spacing $\Delta \nu_{\rm line} =
\hbar/(2 \pi I_3)$, where $I_3$ is the largest moment of inertia,
which is precisely twice as large as the other two equal moments of inertia. Each
line is in fact a ``stack'' from $\sim J_0$ different transitions with equal
frequencies, where $J_0$ is the characteristic total angular momentum quantum
number. We show approximate values of $\Delta \nu_{\rm line}$ for selected symmetric PAHs in
Table \ref{tab:pahs}. Note that this table is neither exhaustive nor
very accurate and only serves the purpose of illustration.

On the other hand, and as we shall illustrate below, a large triaxial molecule has a complex rotational spectrum,
with no closed-form expression, for which the
line spacing is uneven and, more importantly, for which different
transitions do not in general fall at the same frequency. The power is
therefore diluted over a large number of lines with respect to the
symmetric case and decreased by a factor of order $ J_0 \sim 100$ for
the smallest grains, this number increasing with grain size. This
decreased power per line (the lines actually blur into a
quasi-continuum for large molecules), and the additional
complication in the computation of the exact spectrum, concur in making the search for
specific molecules much more difficult if they are triaxial than if
they are symmetric. We must therefore focus our attention on symmetric PAHs if we hope to
achieve any detection through rotational spectroscopy. 

But how symmetric can real \emph{radiating} PAHs be? In order to
radiate, a rotating grain must have a permanent dipole moment. A few PAHs
like corannulene (C$_{20}$H$_{10}$) do possess such a permanent dipole
moment because of their non-planarity. Symmetric non-planar PAHs
nevertheless probably constitute an even smaller fraction of the PAH
family. For planar PAHs, on the other hand, a
permanent dipole moment must come from some imperfection in the grain. Besides the restricted
case where the dipole moment arises from an association with a metal
complex outside of the plane and aligned with the grain's axis of
symmetry, most imperfections leading to a dipole moment
(super-hydrogenation, de-hydrogenation, substitution of a carbon atom
by a nitrogen atom, etc...) will invariably break the symmetry of the
inertia tensor at some level. How exactly this affects the
rotational spectrum is a
quantitative question, which we address in this section. 

Before moving on to the computation, we point out that we shall assume
PAHs to be \emph{rigid} rotors. Laboratory studies of corannulene
indeed show no
significant centrifugal distortion to its rotational spectrum \citep{Lovas_2005,
  Pilleri_2009}. Since small PAHs spend a very large fraction of their
time in their vibrational ground state, we shall moreover assume that they are
described by a single set of rotational constants.

\begin{table*}
\caption{Approximate line spacing $\Delta \nu_{\rm line} = 2 A_3$ for a few symmetric PAHs. They were
  computed assuming a perfectly hexagonal carbon skeleton with a C-C
  bond length of 1.4 \AA \ and a C-H bond length of 1.1 \AA \ for the
  peripheral hydrogen atoms (not shown). Numbers on coronene and
  circumcoronene label the non-equivalent sites for single substitution of a
  carbon atom by a nitrogen atom.} \label{tab:pahs}
  \begin{tabular}{llllll} \hline \hspace{1cm}  Structure &
    \hspace{1cm}   Formula & \hspace{0.5cm} Line spacing (MHz) 
     & \hspace{1cm} Structure &   \hspace{1.3cm}  Formula &  Line spacing (MHz) \\
      \hline 
      \parbox[c]{2em}{\includegraphics[width=1.4in]{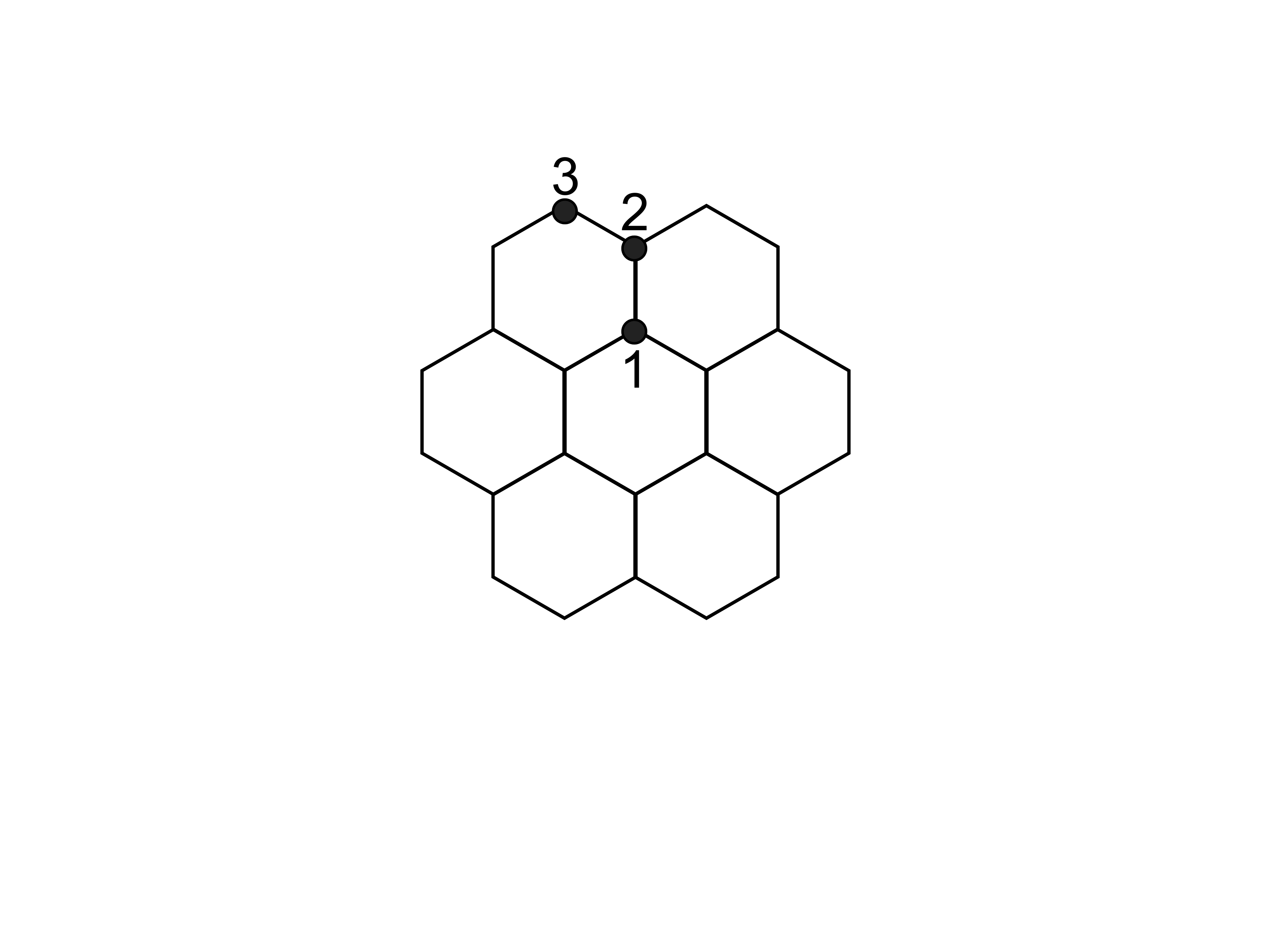}} & 
       \hspace{0.4cm}  \begin{tabular}[c]{@{}c@{}}C$_{24}$H$_{12}$\\(coronene)\end{tabular} &\hspace{0.7cm} 340   & 
      \parbox[c]{2em}{\includegraphics[width=1.4in]{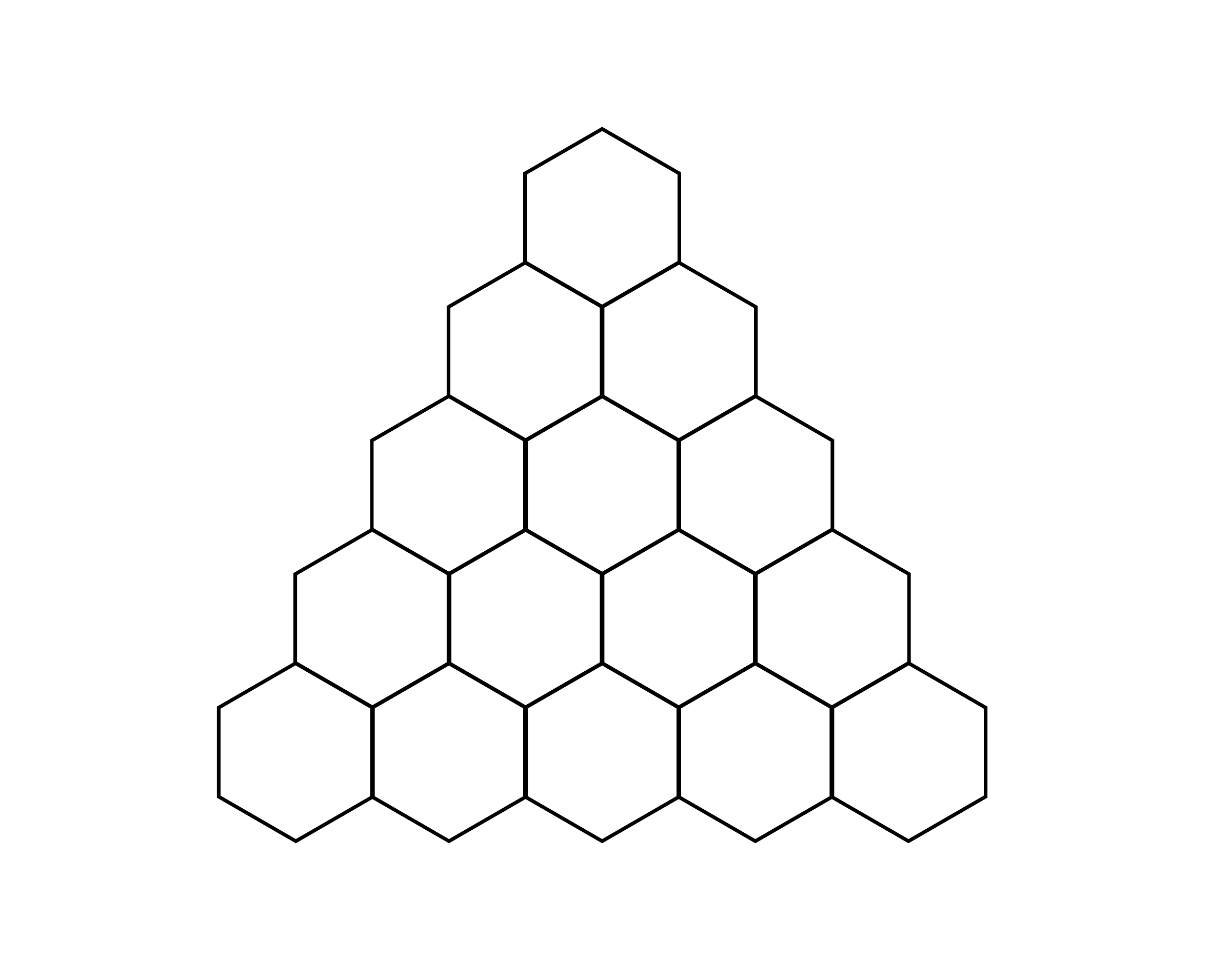}} &\hspace{1.3cm}  C$_{46}$H$_{18}$ & \hspace{0.6cm}  83.0 \\
      \hline 
      \parbox[c]{2em}{\includegraphics[width=1.4in]{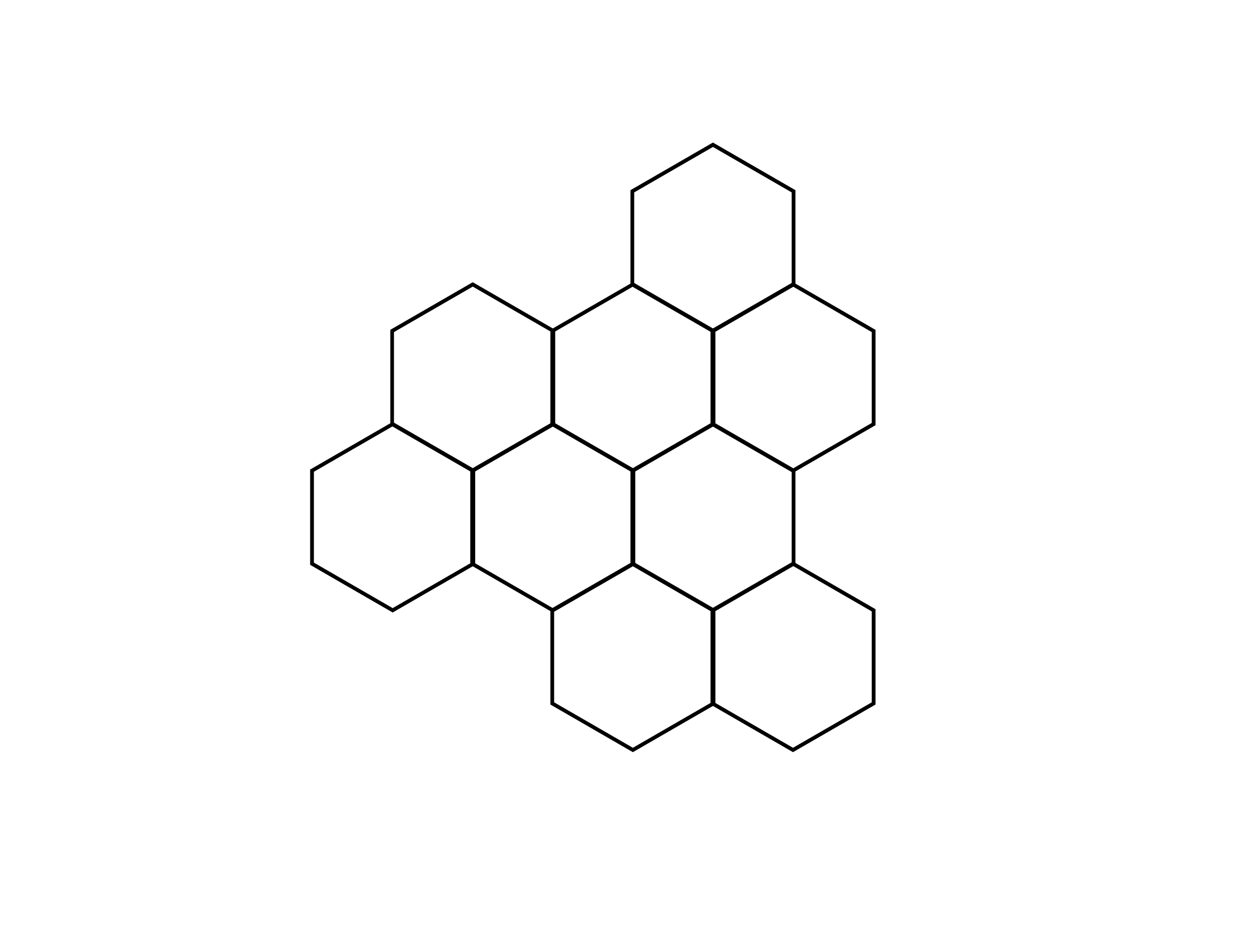}} &   \hspace{1cm}  C$_{31}$H$_{15}$ & \hspace{0.7cm}  192  & 
      \parbox[c]{2em}{\includegraphics[width=1.4in]{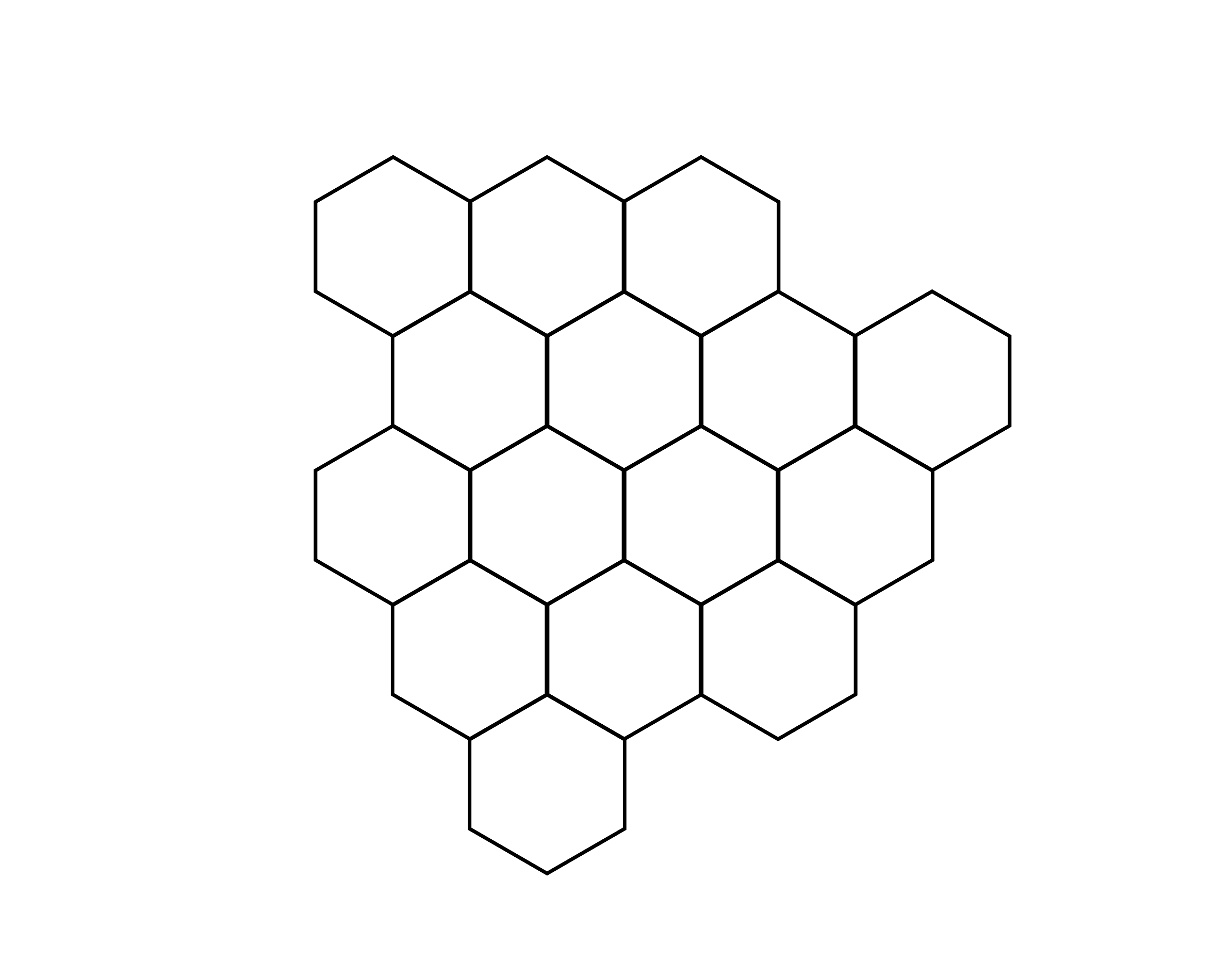}} &  \hspace{1.3cm}  C$_{46}$H$_{18}$ &\hspace{0.6cm} 89.0 \\
      \hline 
      \parbox[c]{2em}{\includegraphics[width=1.4in]{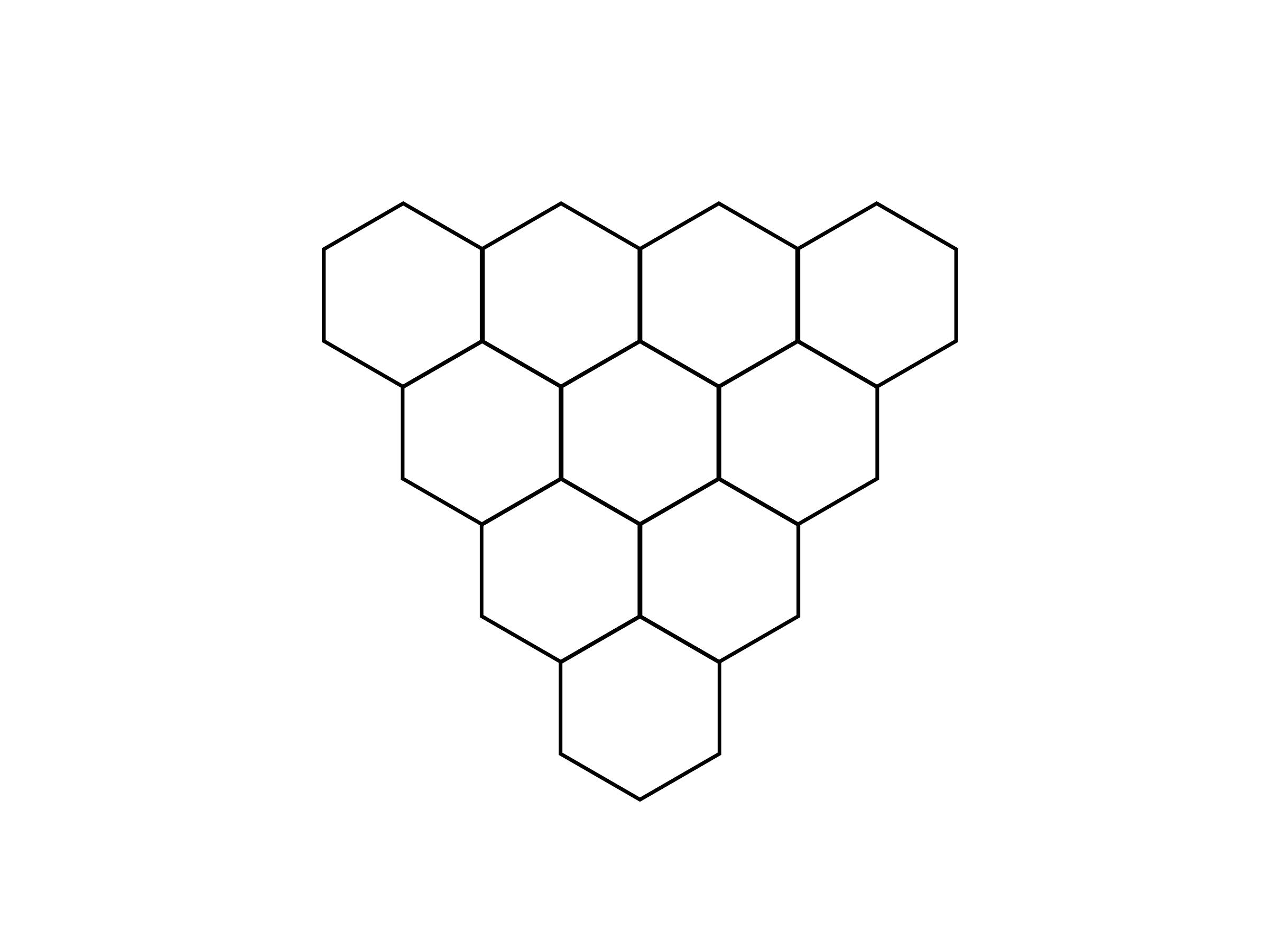}} & \hspace{1cm}  C$_{33}$H$_{15}$ &\hspace{0.7cm}  164 &
      \parbox[c]{2em}{\includegraphics[width=1.4in]{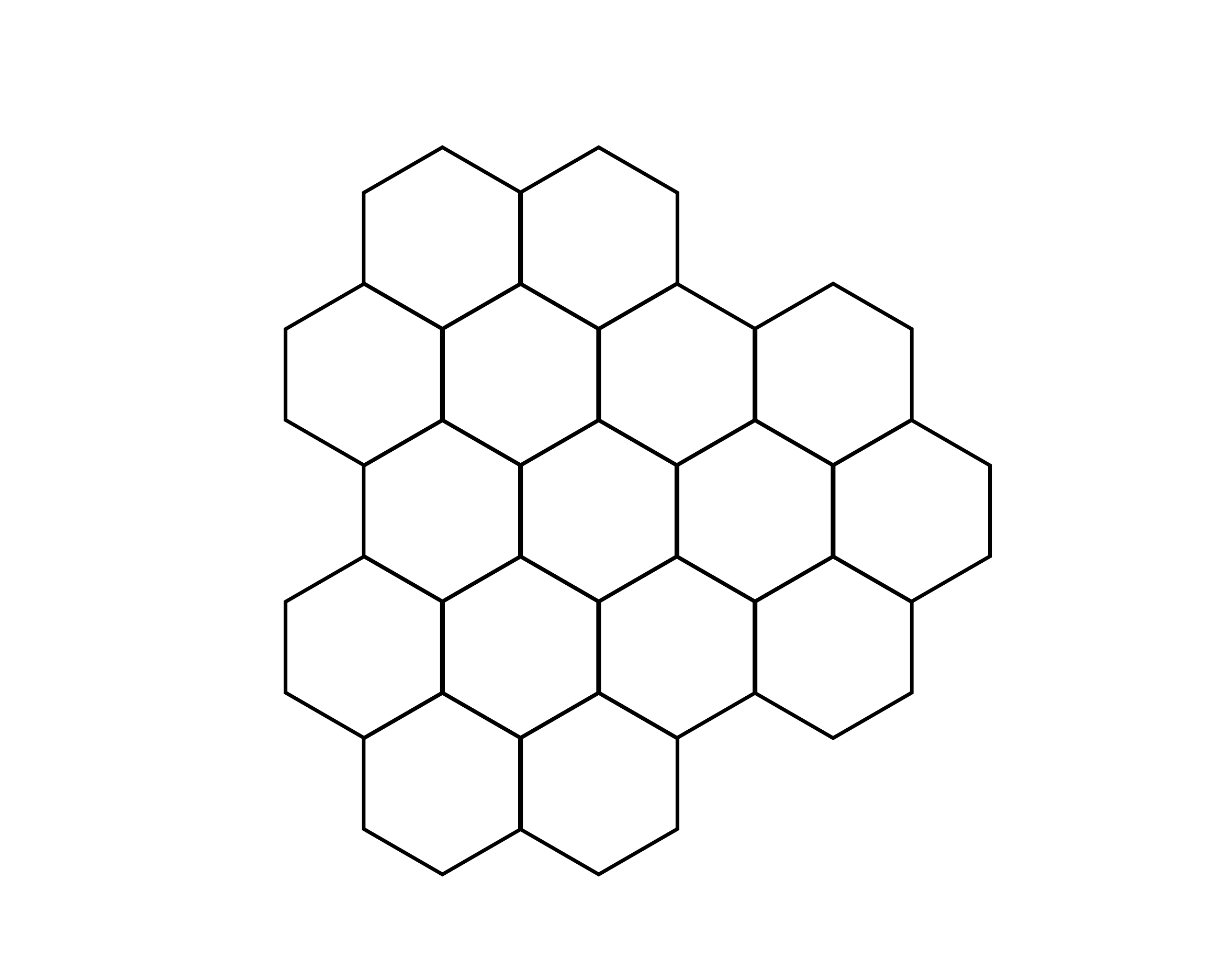}} &  \hspace{1.3cm}  C$_{48}$H$_{18}$ & \hspace{0.6cm}  83.1 \\
      \hline 
      \parbox[c]{2em}{ \includegraphics[width=1.4in]{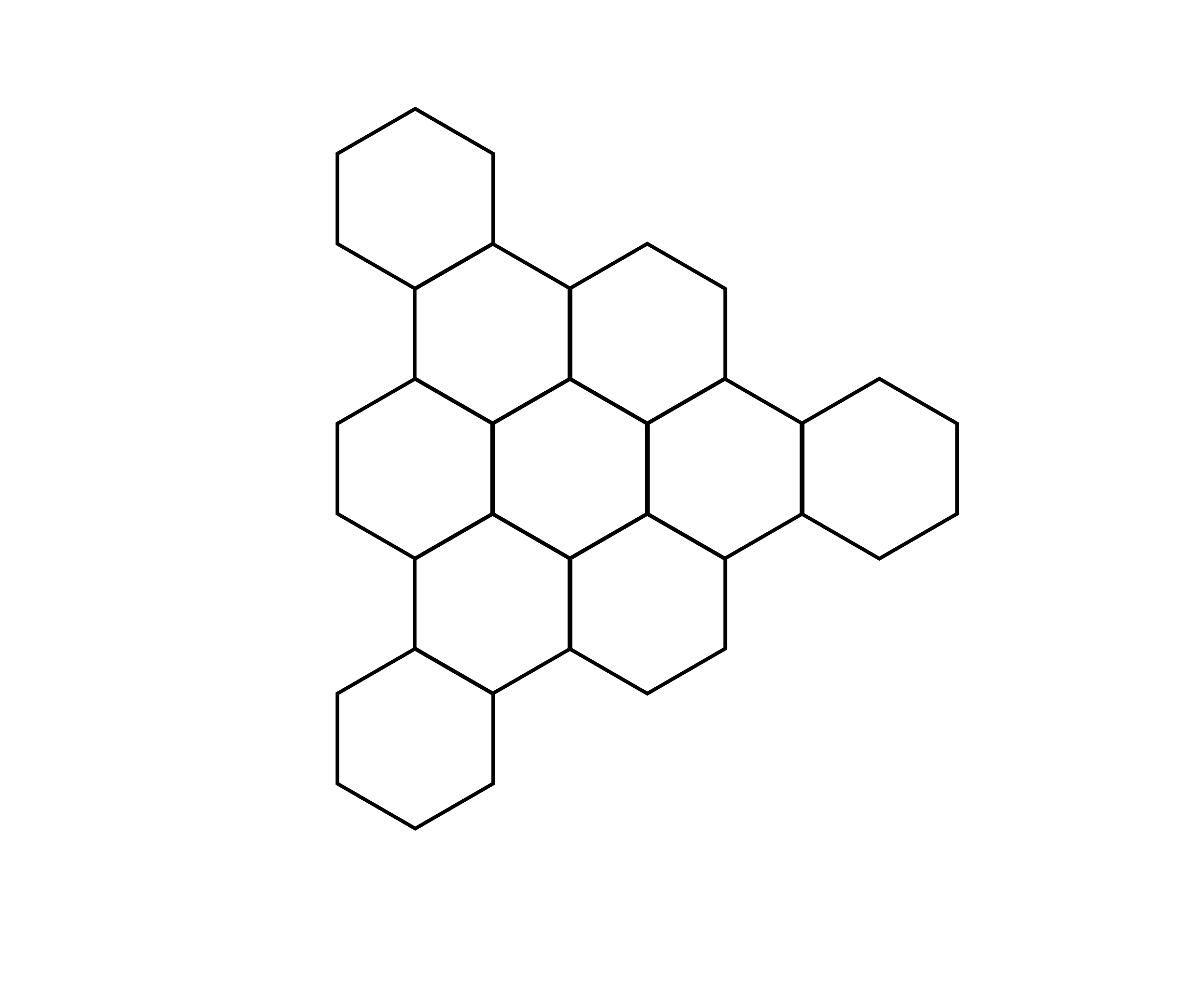}} & \hspace{1cm}  C$_{36}$H$_{18}$ & \hspace{0.7cm}  128 &
      \parbox[c]{2em}{\includegraphics[width=1.4in]{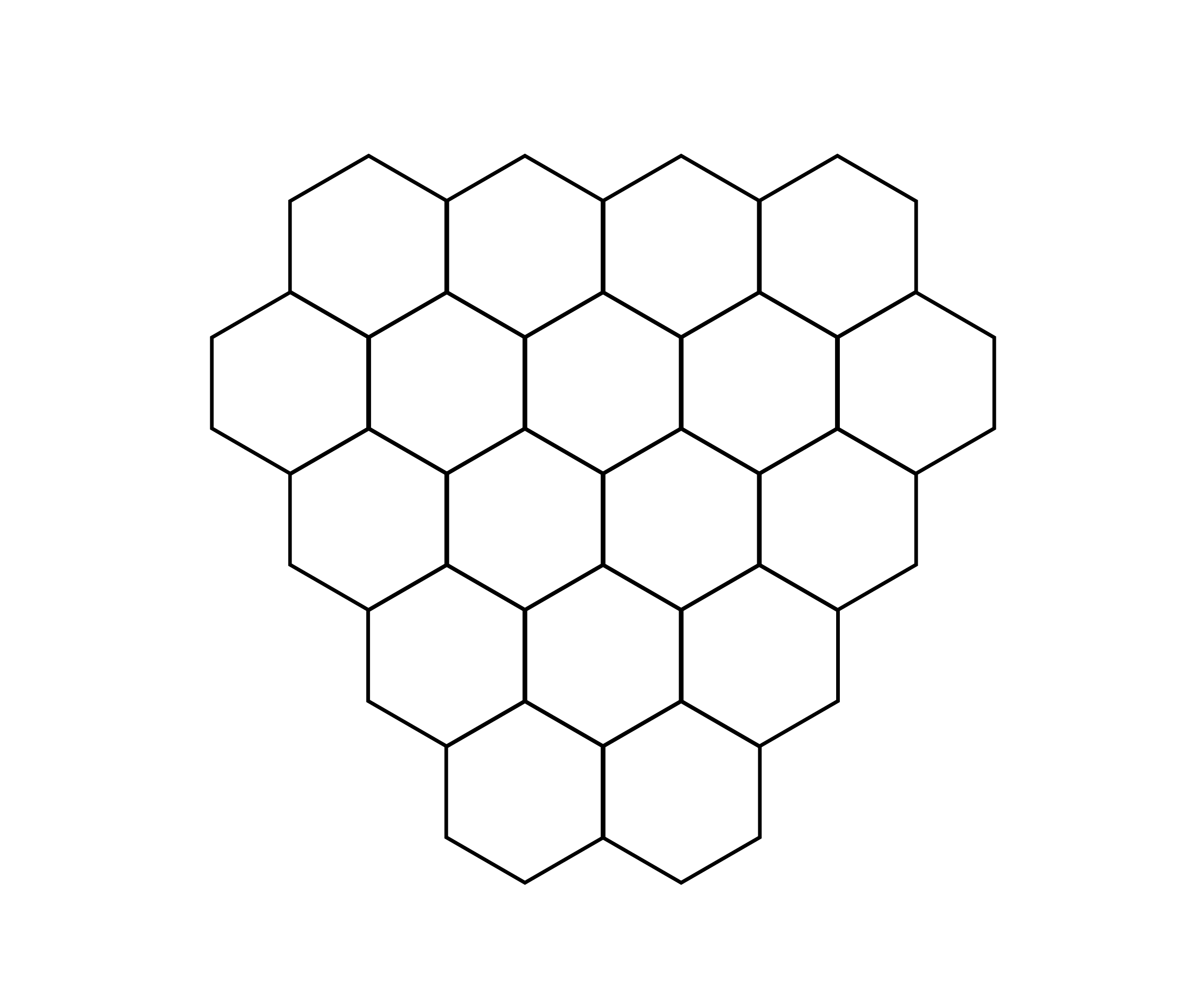}} & \hspace{1.3cm}  C$_{52}$H$_{18}$ & \hspace{0.6cm}  70.7 \\
      \hline 
      \parbox[c]{2em}{\includegraphics[width=1.4in]{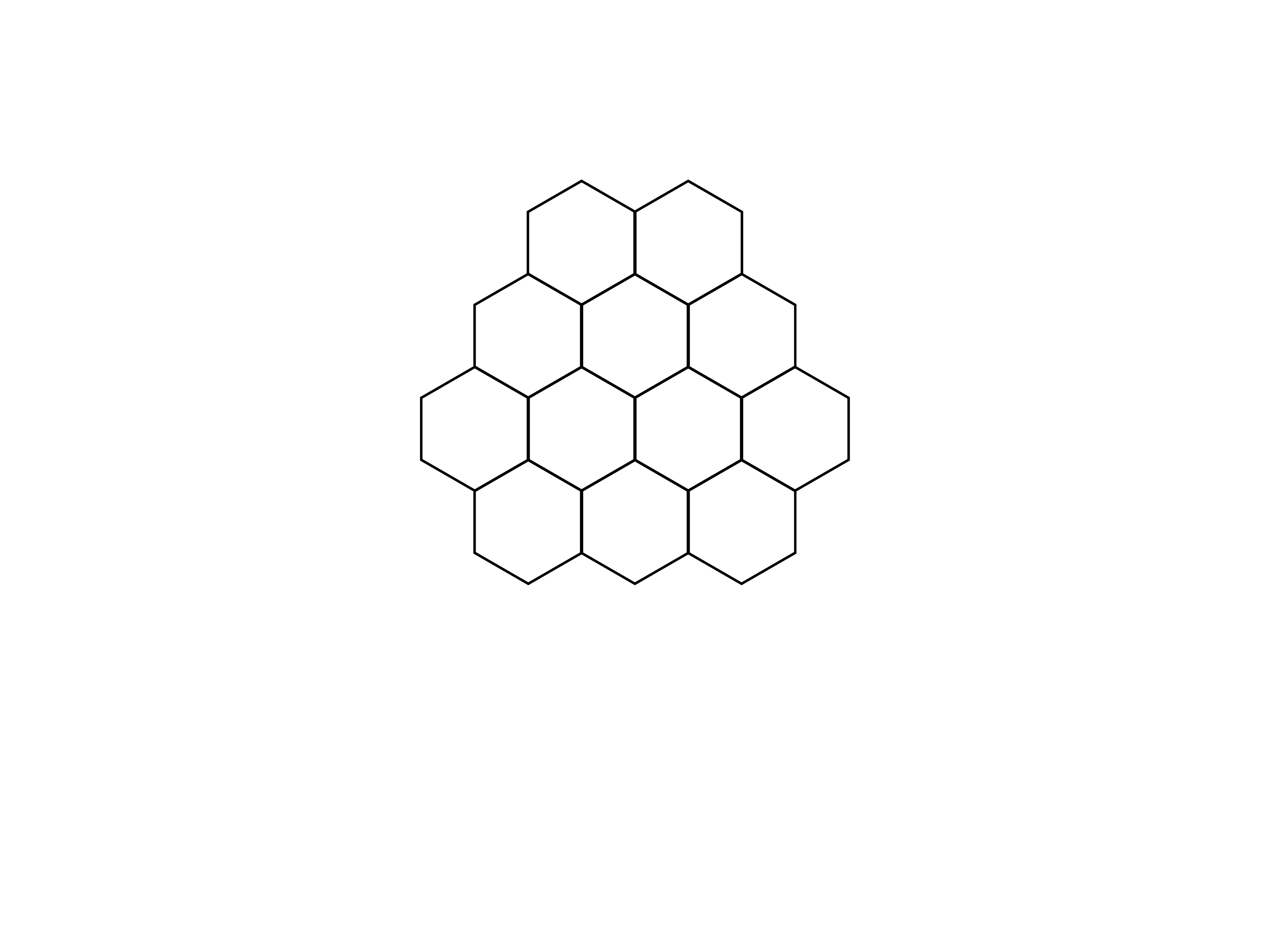}} & \hspace{1cm}  C$_{37}$H$_{15}$ & \hspace{0.7cm}  142 & 
      \parbox[c]{2em}{\includegraphics[width=1.4in]{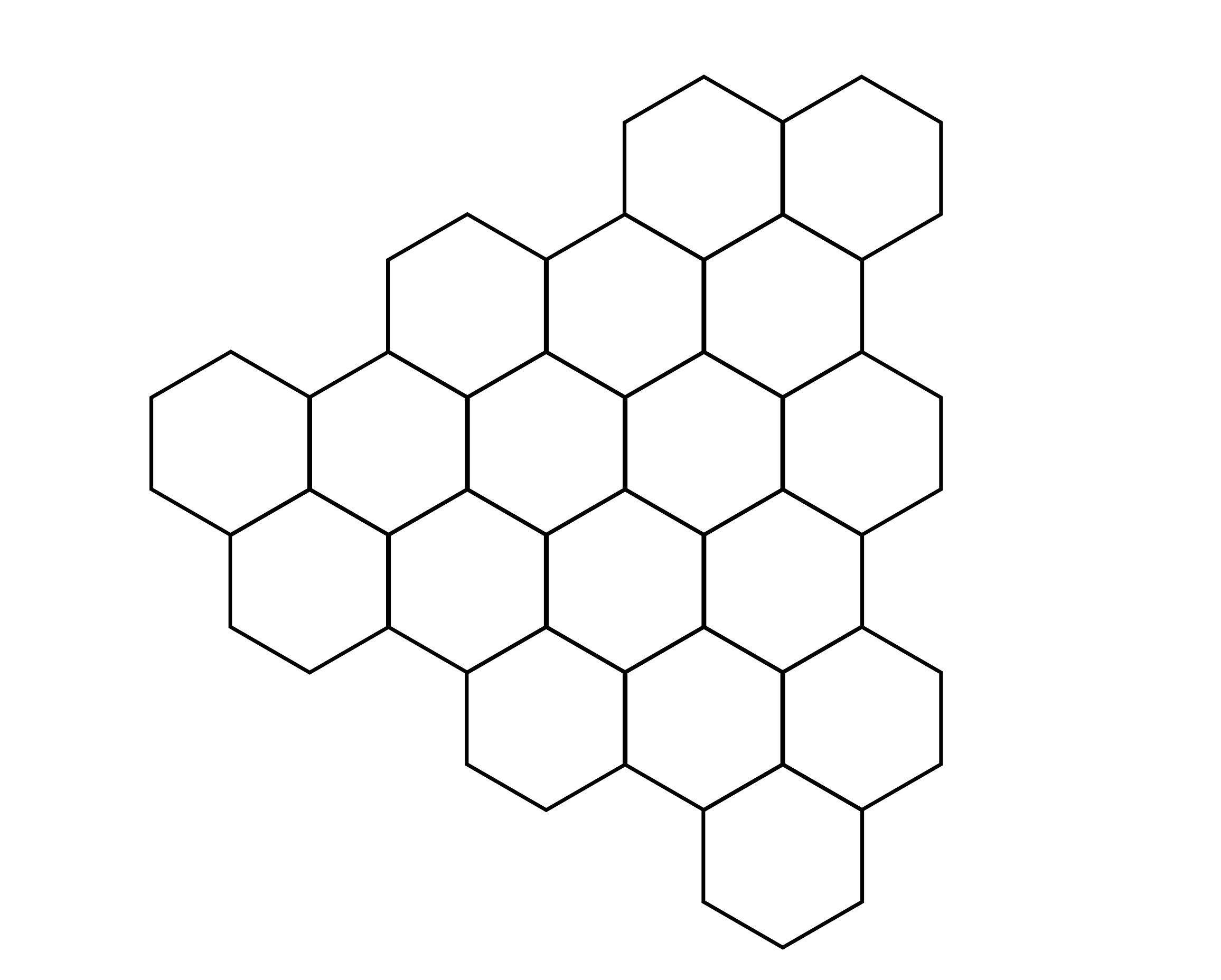}} &  \hspace{1.3cm}  C$_{55}$H$_{21}$ & \hspace{0.6cm} 59.3\\
      \hline 
      \parbox[c]{2em}{\includegraphics[width=1.4in]{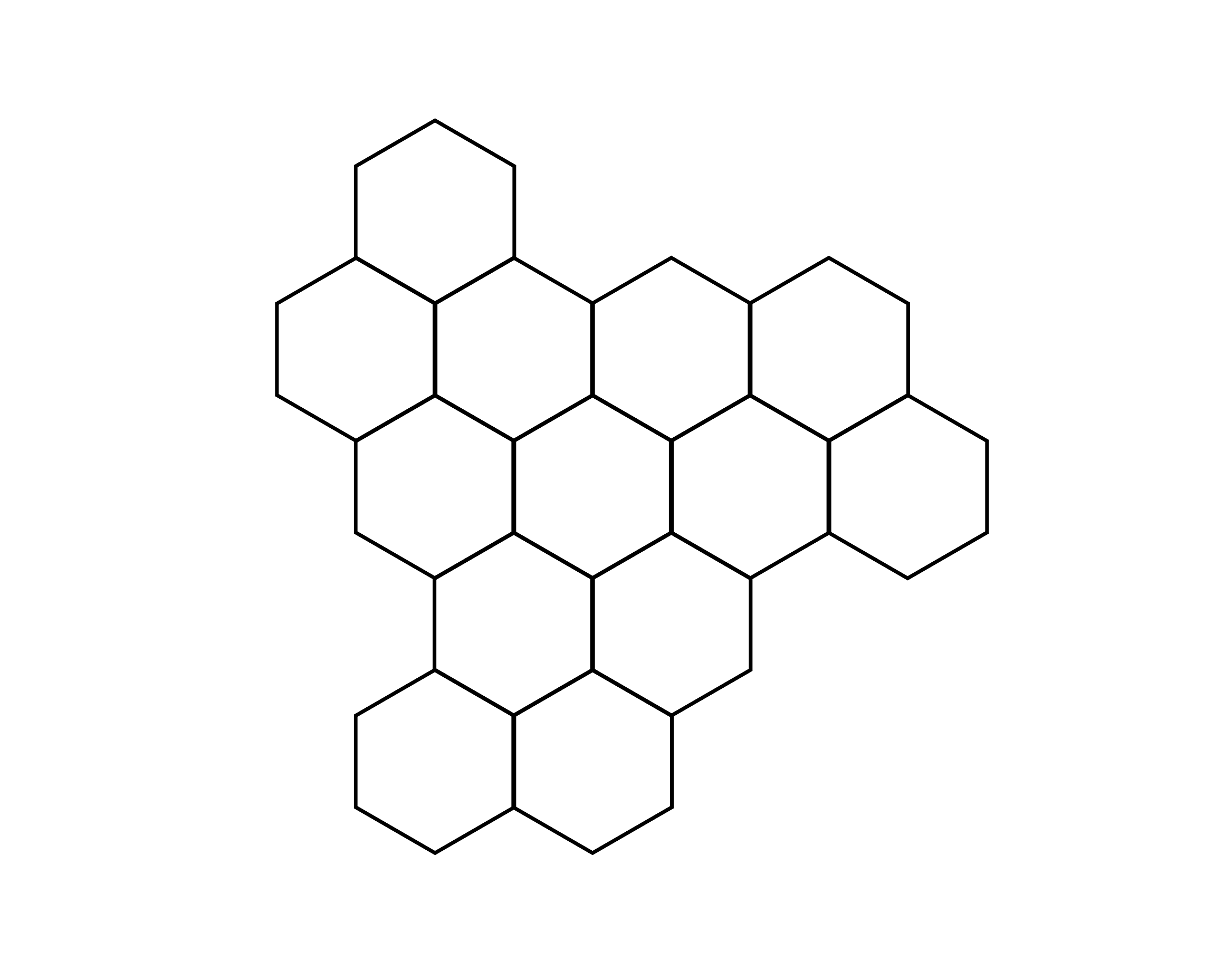}} & \hspace{1cm}  C$_{42}$H$_{18}$ & \hspace{0.7cm} 101 & 
      \parbox[c]{2em}{\includegraphics[width=1.4in]{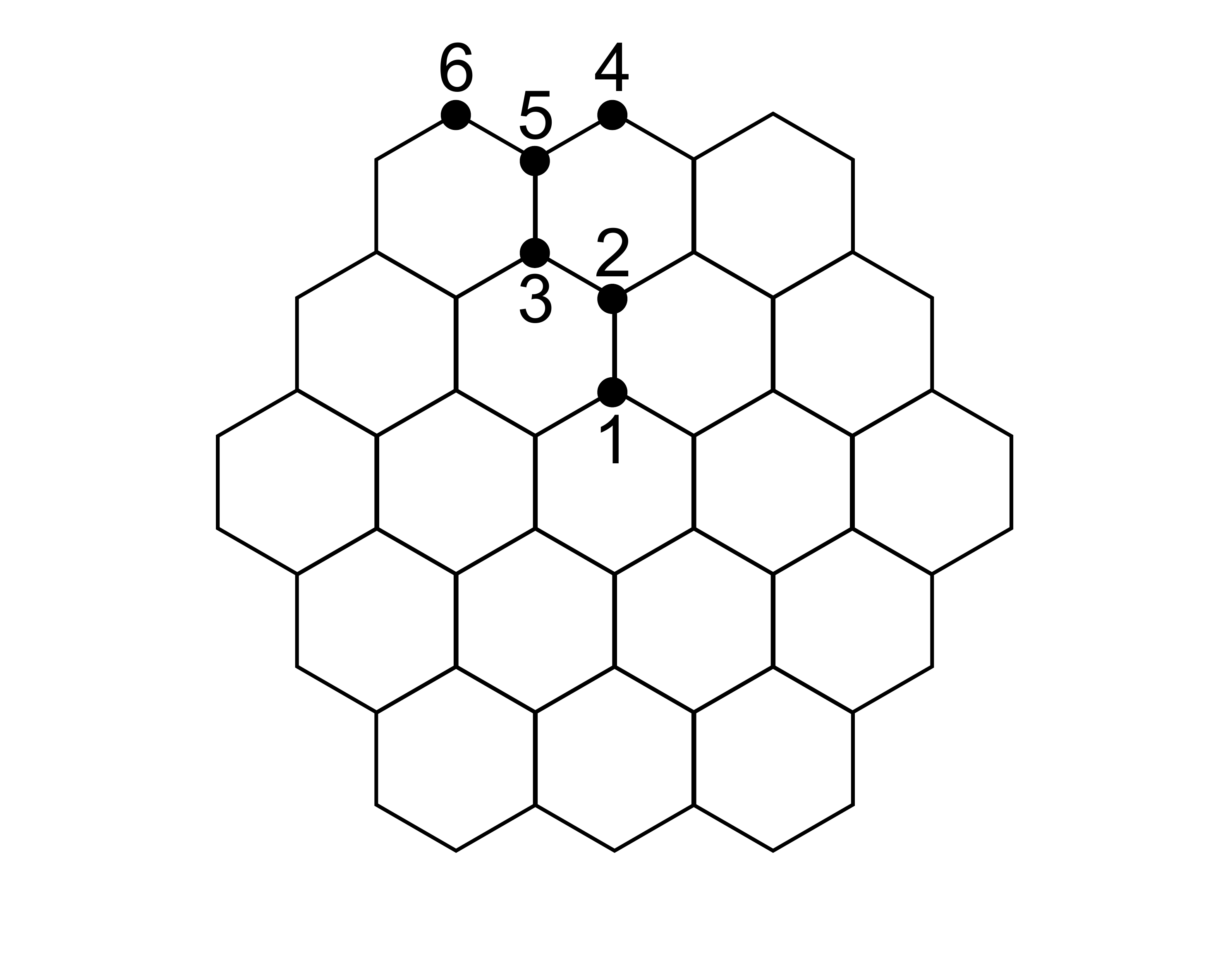}} & 
        \hspace{0.3cm}  \begin{tabular}[c]{@{}c@{}}C$_{54}$H$_{18}$\\(circumcoronene)\end{tabular} & \hspace{0.6cm} 67.6\\
      \hline 
      \parbox[c]{2em}{\includegraphics[width=1.4in]{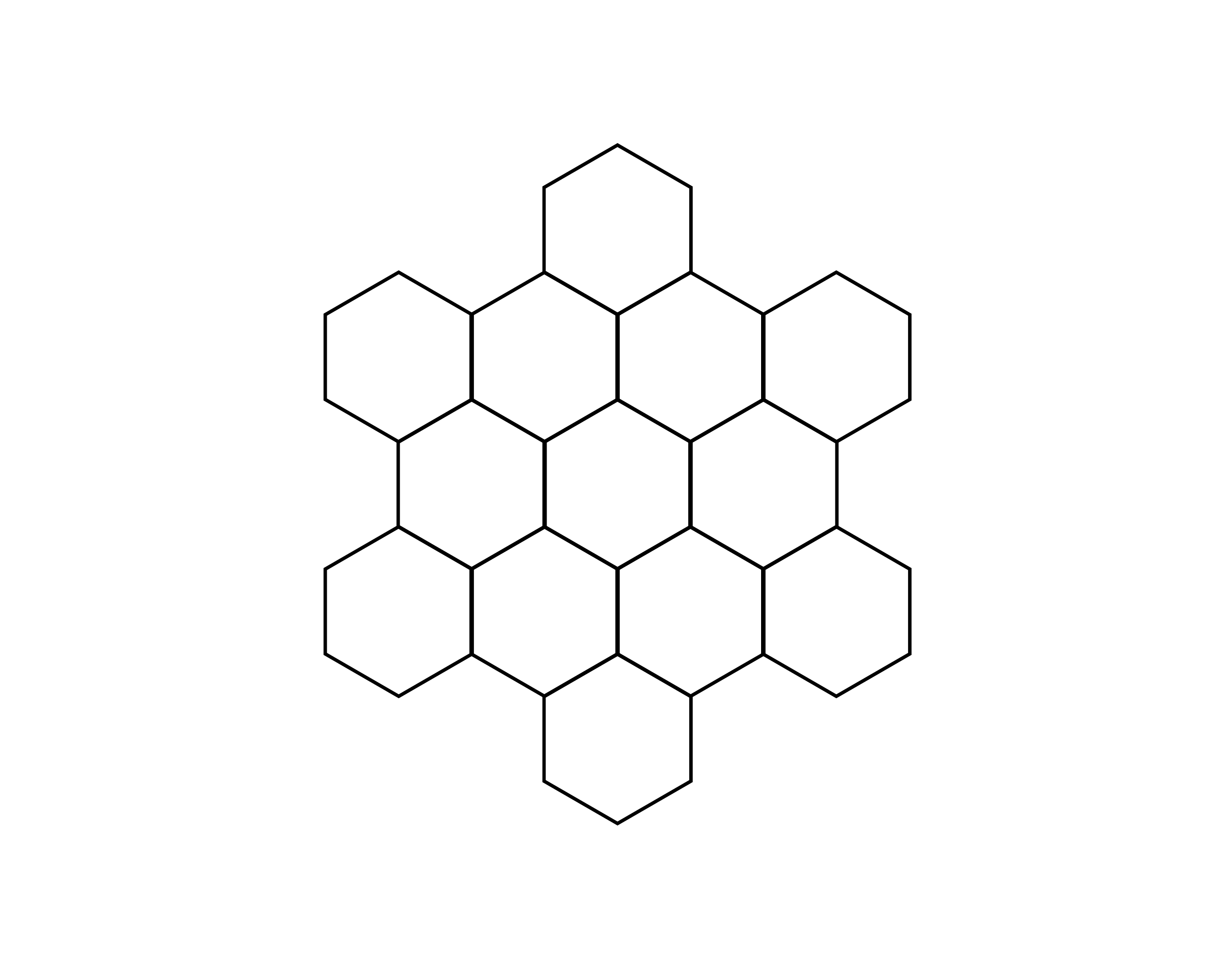}} &  \hspace{1cm}  C$_{42}$H$_{18}$ &\hspace{0.7cm}  108 & 
      \parbox[c]{2em}{\includegraphics[width=1.4in]{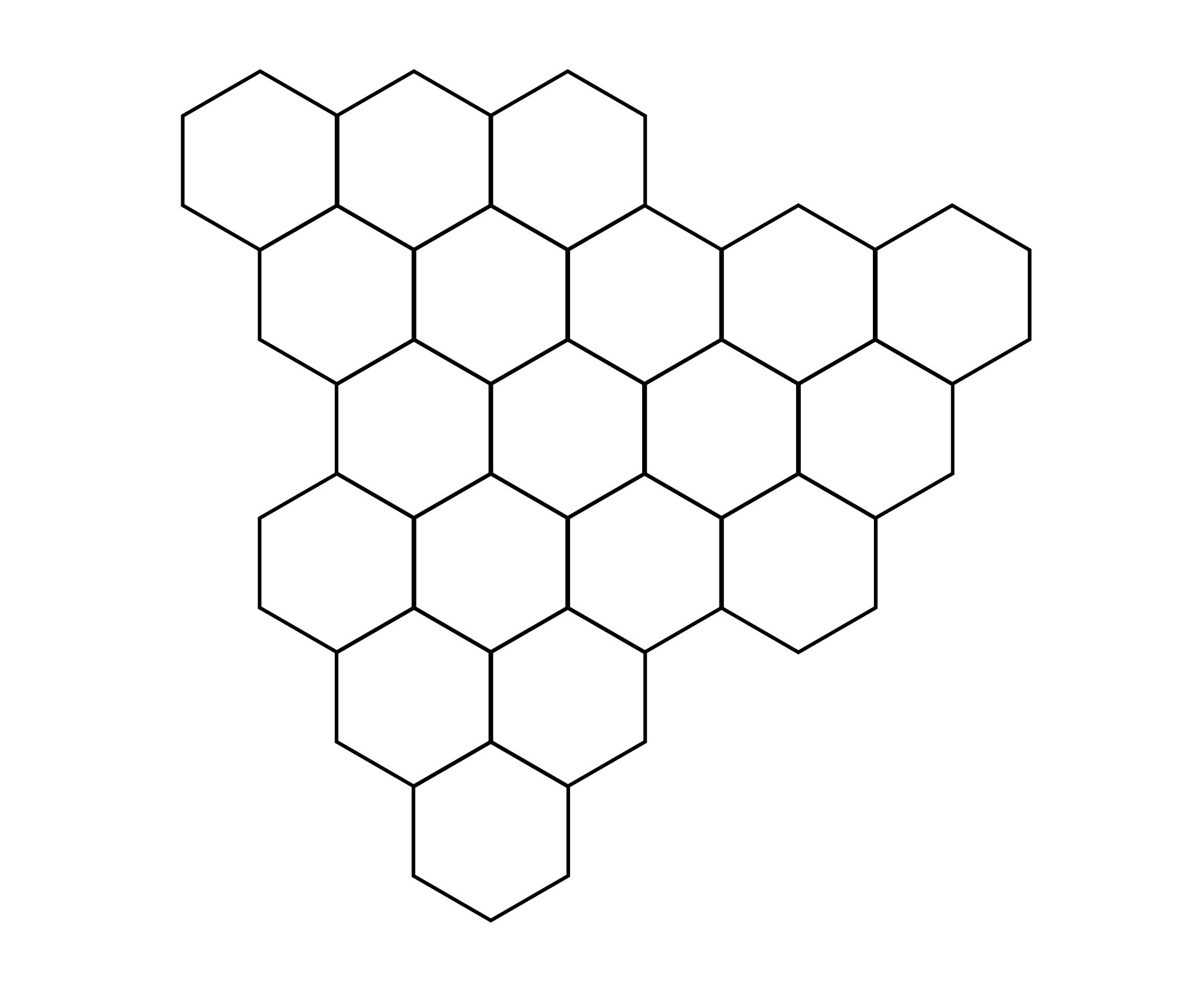}} &\hspace{1.3cm}  C$_{57}$H$_{21}$ & \hspace{0.6cm}  54.8\\
      \hline
  \end{tabular}
\end{table*}

\subsection{Perturbed quantum-mechanical rigid symmetric top}

Computing the transition frequencies and line intensities of a
triaxial quantum-mechanical rotor does not present any major
conceptual difficulty, see for example
\cite{VanWinter_1954}. However, there are no closed form solutions to
this problem. In order to understand the effect of a small degree of
triaxiality analytically, we treat the problem perturbatively, assuming a small
departure from a rigid symmetric-top rotor. This allows us to write
down closed-form expressions for the rotational transitions. 

We consider a rigid triaxial rotor with principal moments of
inertia $I_i$ and angular momentum operators along the principal axes
$L_i$. Its rotational hamiltonian is
\beq
H = \sum_i \frac{L_i^2}{2 I_i} \equiv \frac{2 \pi}{\hbar} \sum_i A_i L_i^2 
\eeq
where $\hbar = h/(2 \pi)$, $h$ being the Planck constant, and the
$A_i$ are the rotational constants, which have the dimension of a frequency:
\beq
A_i \equiv \frac{h}{8 \pi^2  I_i}.
\eeq
Here we shall assume for definiteness $I_1 \leq I_2 < I_3$, i.e. $A_1 \geq A_2 > A_3$. A nearly symmetric oblate top has $A_1
\approx A_2$. Defining
\beq
\overline{A}_{12} \equiv \frac{A_1 + A_2}{2},
\eeq
we rewrite the Hamiltonian in the form
\barr
\frac{\hbar}{2 \pi} H &=& A_3 L_3^2 + \overline{A}_{12}(L_1^2 + L_2^2) +
\frac{A_1 - A_2}{2}(L_1^2 - L_2^2) \nonumber\\
&=& \overline{A}_{12}L^2 -\left(\overline{A}_{12}- A_3\right) L_3^2
+ \frac{A_1 - A_2}{2}(L_1^2 - L_2^2), ~~~~
\earr
where $L^2 \equiv \sum_i L_i^2$. We recognise this expression as the
sum $H = H^0 + V$ of the hamiltonian $H^0$ of a symmetric rotor with
rotational constants $A_1' = A_2' = \overline{A}_{12}$ and $A_3' =
A_3$, and a perturbation $V$ such that 
\beq
\frac{\hbar}{2 \pi}  V\equiv \frac{A_1 - A_2}{2}(L_1^2 - L_2^2)\equiv
\epsilon \left( \overline{A}_{12} - A_3\right) (L_1^2 - L_2^2), 
\eeq
where we have defined our perturbation parameter as in \cite{Townes_1975}:
\beq
\epsilon \equiv \frac{A_1 - A_2}{A_1 + A_2 - 2 A_3}.\label{eq:def:epsilon}
\eeq
The eigenfunctions of the unperturbed hamiltonian $H^0$ are the
wavefunctions $|J K
M\rangle$, with total angular momentum $L^2 =  \hbar^2 J(J+1)$, angular
momentum along the $e_3$ axis $L_3 =  \hbar K$ and angular momentum along
some fixed direction in space $L_z =  \hbar M$, where $J, K, M$ are
integers and $-J \leq K, M \leq J$. The corresponding
unperturbed eigenvalues
are
\beq
E^0_{J K} = h\left[\overline{A}_{12} J(J+1) -\left( \overline{A}_{12}- A_3\right) K^2\right].
\eeq
The perturbation hamiltonian can be rewritten in terms of the operators
$L_{\pm} \equiv L_1 \mp i L_2$ as follows:
\beq
\frac{\hbar}{2 \pi}  V = \frac{\epsilon}{2} \left( \overline{A}_{12} - A_3\right) \left(L_+^2 + L_-^2\right).
\eeq
The operators $L_{\pm}$ are the raising and lowering operators for the
projection of the angular momentum along $e_3$ \citep{VanWinter_1954, Edmonds_1960}:
\beq
L_{\pm} |J K M\rangle = \hbar \left[J(J+1) - K(K \pm 1)\right]^{1/2} |J (K\pm1) M\rangle.
\eeq 
The only non-vanishing matrix elements of the perturbation are therefore
\barr
\langle J (K\pm 2) M |V | J K M \rangle =
h \frac{\epsilon}{2} \left(\overline{A}_{12} - A_3\right) \hspace{3cm} \nonumber\\
\times \left[J(J+1) - K(K \pm 1)\right]^{1/2} \left[J(J+1)  - (K \pm
  1)(K \pm 2)\right]^{1/2}. ~~
\earr
For a given $J$ and $M$, the states $\pm K$ (with $K \neq 0$) are degenerate for the
unperturbed hamiltonian, and
the second-order corrections to the energy levels, $\delta E$,  are given by solving
the secular equation \citep{Landau_QM}:
\beq
\textrm{det}\left[V_{K_1, K_2} + \sum_{K' \neq \pm K}
  \frac{V_{K_1, K'} V_{K', K_2}}{E^0_{JK} -
    E^0_{J K'}} - \delta E \ \delta_{K_1, K_2} \right] = 0,
\eeq
where $K_1, K_2 = \pm K$, $V_{K,K'} \equiv \langle J K M |V| J K' M\rangle$ and the above
equation is to be understood as a 2 by 2 determinant in the subspace
spanned by $|J K M\rangle$ and $|J (-K) M\rangle$.

Let us first consider the subspace spanned by $K = \pm 1$. Since the matrix
element $V_{1, -1}$ is non zero, the energy shift is of first order in
the perturbation parameter. The
proper eigenfunctions are 
\beq
|J 1_{\pm} \rangle \equiv \frac1{\sqrt{2}} \left(|J (+1) M \rangle \pm |
  J (-1) M\rangle \right),
\eeq
which have perturbed energies given by, to lowest order,
\beq
\delta E_{J, 1_{\pm}} = \pm h \frac{\epsilon}{2} \left( \overline{A}_{12}  - A_3\right) J(J+1). \label{eq:dEJ1}
\eeq
We now consider the subspace spanned by $K = \pm 2$. Here the matrix
element $V_{-2, +2}$ does vanish so the perturbation is of second
order. Again, one can show that the eigenfunctions of the perturbed
hamiltonian are
\beq
|J 2_{\pm} \rangle \equiv \frac1{\sqrt{2}} \left(|J (+2) M \rangle \pm |
  J (-2) M\rangle \right),
\eeq
with perturbed energies such that 
\barr
\frac{\delta E_{J, 2+}}{h ( \overline{A}_{12} -
  A_3)} = - \frac{\epsilon^2}{8} \left[ \frac56 J^2(J+1)^2 + J(J+1) - 12\right], \label{eq:dEJ2+}\\
\frac{\delta E_{J, 2-}}{h ( \overline{A}_{12} -
  A_3)} = \frac{\epsilon^2}{8} \left[\frac16 J^2(J+1)^2 - 3J(J+1) + 12\right].\label{eq:dEJ2-}
\earr
Finally, for $|K| \neq 1, 2$, the eigenfunctions $|J K M\rangle$
constitute an appropriate basis for perturbation theory (i.e. the
exact eigenfunctions are only perturbatively different from them), and
the energy shifts are of second order in $\epsilon$ and given by 
\beq
\frac{\delta E_{JK}}{h( \overline{A}_{12} - A_3) } = -  \frac{\epsilon^2}{8} \left[\frac{J^2(J+1)^2}{K^2 -
    1} +2 J(J+1) - 3 K^2 \right]. \label{eq:dEJK}
\eeq
This formula is also valid for $K = 0$.
Splitting of the $\pm K$ degenerate eigenstates only enters through
higher-order corrections for $|K| \neq 1, 2$. We have checked that our expressions (\ref{eq:dEJ1}), (\ref{eq:dEJ2+})
(\ref{eq:dEJ2-}) and (\ref{eq:dEJK}) coincide with exact expressions
for the energy levels of an asymmetric top computed for low
$J$-values, when Taylor-expanded in the asymmetry parameter. Such expressions
can be found in Chapter 4 of \cite{Townes_1975}, for example.

For large $J$, and $K \sim J$ this expression is of order
$\delta E_{JK} \sim \epsilon^2 E_{JK}^0$, as one would expect. 
For $1 \ll |K| \ll J$, using $A_1 \approx A_2 \approx 2 A_3$
for a planar grain, we
have, to lowest order,
\beq
\frac{\delta E_{JK}}{E^0_{JK}} \approx -\frac{\epsilon^2}{16} \frac{J^2}{K^2}.
\eeq
This means that perturbation theory is valid only as long as $K \gg
(\epsilon/4) J$. In practice, we will consider typically $\epsilon
\sim 10^{-3}- 10^{-2}$ and $J \sim
10^2-10^3$ so the perturbation theory results might only break down for the
smallest few values of $K$ and quickly becomes very accurate as
$|K|$ increases. 

\subsection{Application to a planar grain}

We now specialise to a \emph{planar} grain. For a classical rigid planar
object with normal along $e_3$, the moments of inertia are related by
$I_3 = I_1 + I_2$. However, this relation is not exact for a
quantum-mechanical object due to zero-point fluctuations and
vibration-rotation interactions (i.e. not perfect rigidity). We denote by $\delta$ the dimensionless
inertial defect quantifying the departure from this relation:
\beq
\delta \equiv \frac{I_3 - I_1 - I_2}{I_3}.
\eeq
This quantity can have either sign and is small, $|\delta| \ll 1$.

With this definition and that of $\epsilon$,
(Eq.~\ref{eq:def:epsilon}), we may rewrite $A_1$ and $A_2$, to second
order in $\epsilon$ and first order in $\delta$, in terms of $A_3$:
\barr
A_1 &\approx& \left[2 + \epsilon + \frac12 \epsilon^2 \right] \frac{A_3}{1 - \delta},\\
A_2 &\approx&  \left[2 - \epsilon + \frac12 \epsilon^2\right] \frac{A_3}{1 - \delta}
\earr
We then re-express the unperturbed energy in the following form, to lowest
order in $\epsilon$ and $\delta$:
\beq
\frac{E^0_{JK}}{h A_3} =  2 J(J+1) - K^2 +
\left(\frac{\epsilon^2}{2}+ 2 \delta\right)\left(J(J+1) - K^2\right).
\eeq
This allows us to rewrite the final energy values as follows, to lowest
order in $\epsilon$, and for $|K| \neq 1, 2$:
\barr
\frac{E_{JK}}{h A_3} &=&  2 J(J+1) - K^2 + 2 \delta  \left(J(J+1) - K^2\right)\nonumber\\
&-&\frac{\epsilon^2}{8}\left(\frac{J^2(J+1)^2}{K^2-1} - 2 J(J+1) +
  K^2\right). \label{eq:EJK-final}
\earr
We have checked that this expression agrees very well with numerical
computations for slightly asymmetric tops using the online tool
\textsc{pgopher}\footnote{http://pgopher.chm.bris.ac.uk/index.html}
with $J \leq 10$.

\subsection{Emission spectrum}
Since our expressions are most inaccurate for low values of $|K|$ we shall in what follows only consider transitions involving $|K| \geq
3$, and disregard other transitions. Only a small fraction of the
radiated energy is emitted in transitions involving $|K| \leq 2$ for large $J$, and
they therefore only contribute a few weak lines that do not affect
observations. This also allows us to easily compute transition rates,
since the eigenvalues for $|K| > 2$ are the usual $|J K M\rangle$ with
small perturbations.

For the unperturbed symmetric-top, there are four allowed transitions from any
given state $(J,K)$, radiating a power $P$ in the limit\footnote{for exact
  expressions, see for example \cite{thesis}} $J, |K| \gg 1$ given by:
\barr
(J, K) \rightarrow (J-1, K),  && P
= \frac{2 \mu_{\rm op}^2}{3 c^3} (2 \pi \nu)^4 \left(1 -
  \frac{K^2}{J^2}\right),~~~~~~~\label{eq:J-1}\\
(J,|K|) \rightarrow (J, |K|+1),  && P =
\frac{\mu_{\rm ip}^2}{3 c^3}  (2 \pi \nu)^4 \left(1 -
  \frac{K^2}{J^2}\right),~~~~~~~\label{eq:Kpm1}\\
(J,K) \rightarrow (J-1, K\pm1),  && P =
\frac{ \mu_{\rm ip}^2 }{6 c^3}  (2 \pi \nu)^4\left(1 \mp
  \frac{K}{J}\right)^2, ~~~~~~~\label{eq:J-1Kpm1}
\earr
where $\nu$ is the frequency of the transition, $\mu_{\rm op}$ is the out-of-plane component of the permanent
dipole moment (along $e_3$) and $\mu_{\rm ip}$ is its in-plane component. 

We will focus our attention on planar substituted PAHs, for which the
dipole moment is purely in-plane $(\mu_{\rm op} = 0$) and the first transition above is not
allowed (but there is of course no additional
difficulty in considering general orientations of the dipole
moment). Also, the $\Delta J = 0$ transition radiates 174 times less
power than the $\Delta J = -1, \Delta K = \pm 1$ transition if $K$ is
evenly distributed in $[-J, J]$ \citep{thesis}, and radiates no power at all for grains rotating primarily
about their axis of greatest inertia. So we only need to focus on the
third transition above, Eq.~(\ref{eq:J-1Kpm1}). Finally, the perturbed
energy levels only depend on the absolute value of $K$, and the
transition $J, K \rightarrow J-1, K+1$ and $J, K \rightarrow J-1,
-K-1$ have the same frequency and rate. We therefore only need to
account for the $\Delta J = -1, \Delta K = -1$ transitions, provided
we double the power given in Eq.~(\ref{eq:J-1Kpm1}).

\subsection{Impact of defects on the spectrum}

A small asymmetry will have two effects. First, it perturbs the
eigenfunctions to first order in $\epsilon$. This leads to changes to
the rates of the allowed transitions at order $\epsilon$, and to
additional transitions with rates proportional to $\epsilon^2$. These
changes are essentially unobservable. The second and most important
effect is the change to the transition frequencies. The $\Delta J = -1, \Delta K = -1$ transitions
have the following frequencies, to lowest order in $\epsilon$ and $\delta$: 
\barr
&&\nu_{J,K} \equiv \frac{E_{JK} - E_{J-1, K-1}}{h} = A_3 (4 J -2 K +
1)\nonumber\\
 &&+ A_3\frac{\epsilon^2}{4} \left(2 J- K +\frac{J^3}{K^3}(J - 2
    K) \right) + 4A_3\delta ( J - K),~~~~~\label{eq:omegaJK}
\earr
where we have Taylor-expanded the perturbation parts in $J, |K| \gg
1$, i.e. we are neglecting corrections of order $\epsilon^2 A_3,
\delta A_3$ to the transition frequency (typically, if $A_3 \sim 100$
MHz and $\delta \sim \epsilon^2 \lesssim 10^{-4}$, these corrections
are of order 0.01 MHz, much smaller than the turbulent velocity
broadening). Note that for our numerical computations we use the exact
expressions for the energies and transition frequencies.

Let us consider the family of transitions with $K = 2J - J_0$ for
$\frac13 J_0 \leq J \leq J_0$. Their frequencies are
\barr
\nu_{J_0}(J) &\equiv& \nu_{J, 2J - J_0} = A_3 (2 J_0 + 1)\nonumber \\
 &+& A_3\frac{\epsilon^2}{4} (J_0 - J)^3 \frac{3 J - J_0}{(2 J -
     J_0)^3} + 4 A_3 \delta(J_0 - J).
\earr
Note that this expression is not valid near $J = J_0/2$ i.e. $K = 0$,
since we have derived in it the limit $|K| \gg 1$.

From this expression we can draw several consequences:

$(i)$ In the absence of asymmetry and inertial defect, the above transitions all
fall at the same frequency, and appear as a single strong
line, which is really a ``stack''. These stacks of
lines are themselves evenly spaced; their transition energies are a
constant times half an integer: $\nu_{J_0} \equiv A_3(2 J_0 + 1)$. The
rotational spectrum appears as a ``comb'' with constant spacing
$\Delta \nu_{\rm line} = 2 A_3$.

\begin{figure*}
\includegraphics[width = 175 mm]{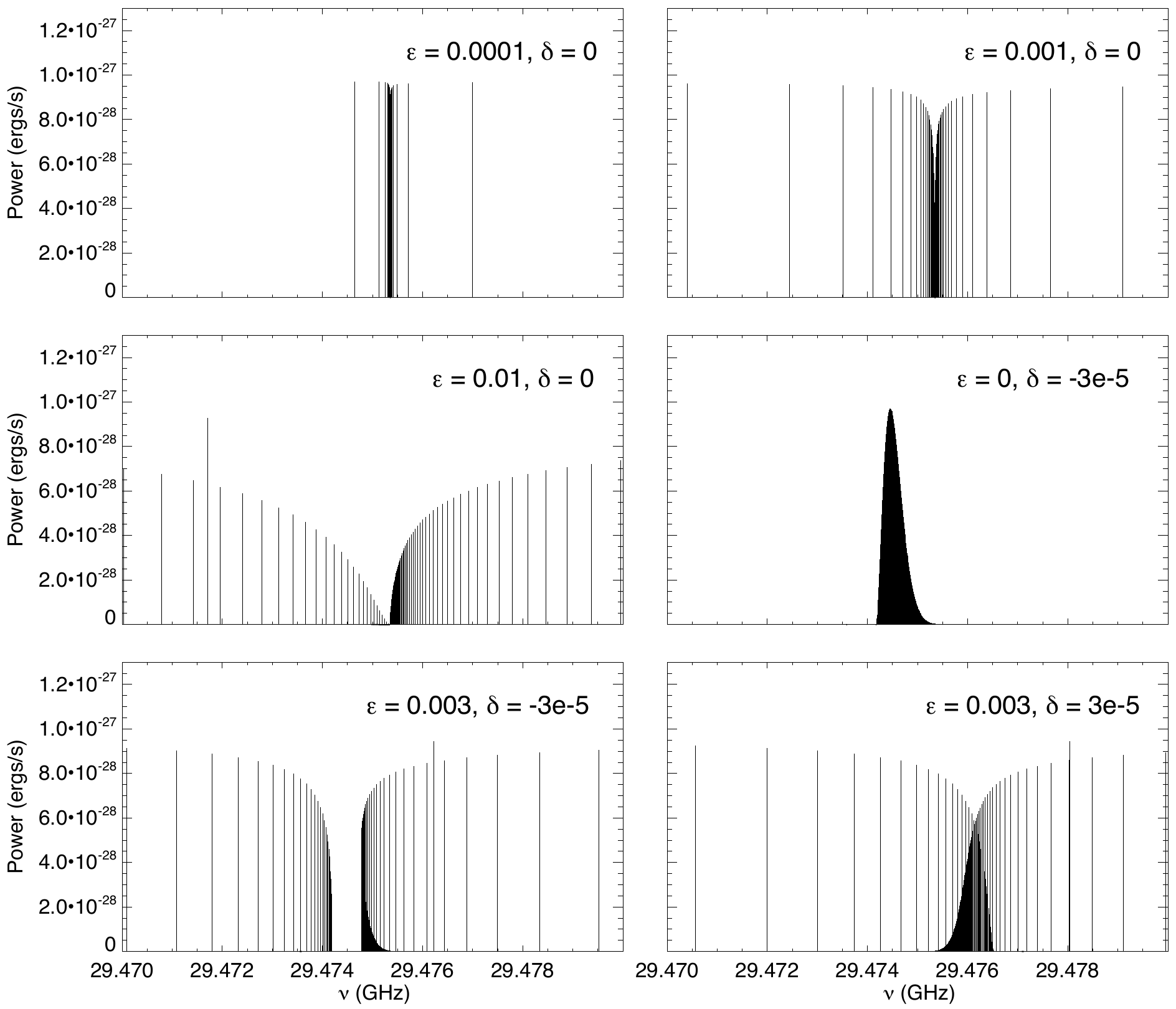}
\includegraphics[width = 175 mm]{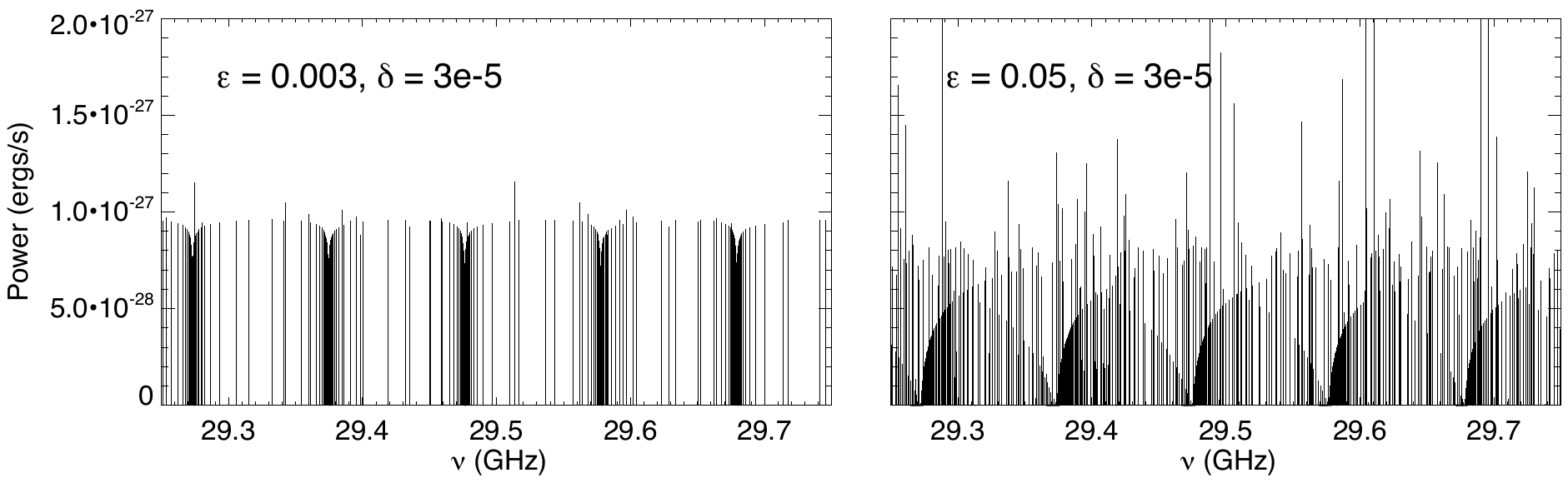}
\caption{Effect of a small asymmetry $\epsilon$ and a small inertial
  defect $\delta$ on the rotational spectrum of a nearly symmetric planar PAH of
  about 40 carbon atoms (corresponding to a line spacing of about 100
  MHz), with dipole moment $\mu_{\rm ip} = 2$ Debye. The probability distribution for the
rotation state was set to $P(J) \propto J^2 \exp[-(J/90)^2]$ and the distribution
$P(K|J)$ was conservatively assumed to be uniform (this is a
physically motivated assumption, see for example \citealt{silsbee}; it
is conservative because a distribution more sharply peaked around $K =
J$ would lead to less spread in the lines). With this distribution
function, the rotational emission peaks around $30$ GHz. The top 6 panels show the detail of clustering of the
  $\Delta J = -1, \Delta K = -1$ transitions with constant $2J - K$
  over a 10 MHz region. For a perfectly symmetric grain with no inertial
  defect all the lines would be stacked at the same frequency. The bottom two panels show the line intensity over a 500
  MHz region. For a small asymmetry ($\epsilon = 0.003$), lines
accumulate at evenly spaced frequencies, whereas for a relatively large
asymmetry ($\epsilon = 0.05$), the spectrum becomes a dense forest
of lines.} \label{fig:lines} 
\end{figure*}

\begin{figure*}
\includegraphics[width = 180 mm]{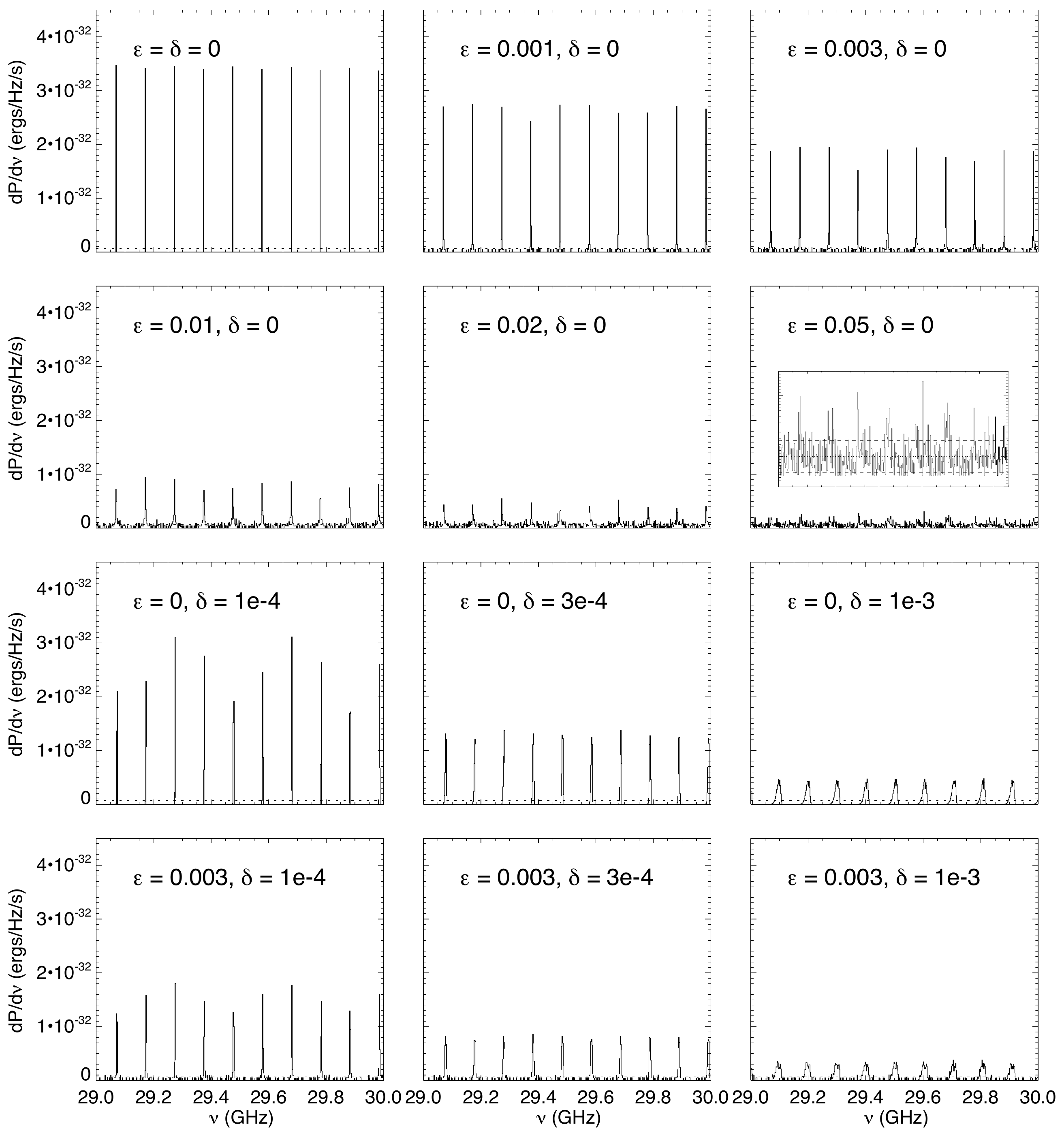}
\caption{Appearance of the rotational spectrum of a PAH
  with line spacing of about 100 MHz observed with a resolution $\Delta
  \nu_{\rm res} = 2$ MHz, for several values of the asymmetry parameter and inertial
  defect. The dotted lines denote the mean. Negative inertial defects produce the same qualitative
  behaviour. The inset in the spectrum for $\epsilon = 0.05, \delta =
  0$ has a smaller $y$-scale and shows that for such a level of asymmetry the line intensity is
  comparable to the intrinsic variance of the spectrum (the dotted
  line denotes the mean and the dashed lines denote the 1-$\sigma$ rms
  fluctuation about the mean).} \label{fig:spectra} 
\hspace{10pt}
\end{figure*}

$(ii)$ The effect of the asymmetry alone is to ``unfold'' each stack of
lines into two branches that accumulate near the central frequency
$\nu_{J_0, J_0}= \nu_{J_0}$, as the correction term vanishes for $ J =
J_0 = K$ and $J = J_0/3 = -K$, see top panels of Fig.~\ref{fig:lines}. The two branches correspond to $0 < K \leq J$
(positive frequency offset) and $-J \leq K < 0$ (negative frequency
offset). The characteristic spread of the lines, obtained for example by
considering the range $-J/2 \leq K \leq J/2$ (corresponding to
$\frac25 J_0 \leq J \leq \frac23 J_0$) is 
\beq
\Delta \nu \approx \frac45 \epsilon^2 \nu_{J_0}.
\eeq
Therefore the bulk of the lines remain within the same resolution bin
of width $\Delta \nu_{\rm res}$ and the observed aspect of the spectrum is roughly unchanged from the case
$\epsilon = 0$ as long as 
\beq
\epsilon \lesssim \sqrt{\frac{\Delta \nu_{\rm
  res}}{\nu}} \approx 10^{-2} \sqrt{\frac{\Delta \nu_{\rm
  res}}{3 ~ \rm MHz} \frac{30~ \rm GHz}{\nu}}, \label{eq:eps_comb}
\eeq
For larger asymmetries, the ``comb'' appearance is preserved but the
contrast of the lines decreases with increasing $\epsilon$. For asymmetries greater than 
\beq
\epsilon_{\rm crit} \approx \sqrt{\frac{\Delta \nu_{\rm line}}{\nu}}
\approx 0.06 \sqrt{\frac{\Delta \nu_{\rm line}}{100 ~\rm MHz} \frac{30
    ~\rm GHz}{\nu}}, \label{eq:eps_crit}
\eeq
each stack of lines is spread over an interval as large as the
characteristic line spacing; the spectrum becomes a noisy
quasi-continuum essentially unusable for identification
purposes. These statements are illustrated in Figs.~\ref{fig:lines}
and \ref{fig:spectra}.

$(iii)$ The effect of the inertial defect alone is to spread the lines
of each stack between $\nu_{J_0}$ and
$\nu_{J_0}(1 + \frac43 \delta)$ (see Fig.~\ref{fig:lines}). Whereas a small asymmetry could still shift part of the
lines very far from their initial frequency (in the limit $|K| \ll J$), an inertial defect spreads the lines over a finite
interval of order $\Delta \nu/\nu \approx \delta$. As long as 
\beq
\delta \lesssim \frac{\Delta \nu_{\rm res}}{\nu} \approx 10^{-4} \frac{\Delta \nu_{\rm
  res}}{3 ~ \rm MHz} \frac{30~ \rm GHz}{\nu},
\eeq
each stack of lines remains clustered within the same resolution
element and the observed spectrum is nearly identical to that of a
perfectly planar grain. For larger values the spectrum gets spread over
a few resolution bins, but still conserves a periodic structure. Only
when $\delta$ becomes comparable to the critical value 
\beq
\delta_{\rm crit} \approx \frac{\Delta \nu_{\rm line}}{\nu} \approx 0.003 \frac{\Delta \nu_{\rm line}}{100 ~\rm MHz} \frac{30
    ~\rm GHz}{\nu}
\eeq
does the periodic structure get effectively smeared out. 

We illustrate the effect of a small asymmetry or inertial defect on
the distribution of lines in Fig.~\ref{fig:lines} and on the resulting observed spectrum in
Fig.~\ref{fig:spectra}.

To summarise, this analysis shows that for an asymmetry $\epsilon$ less than a few
percent and a dimensionless inertial defect $\delta$ less than a few times $10^{-4}$, the rotational
spectrum conserves the appearance of a ``comb'' when observed with a
few MHz resolution. Higher asymmetries rapidly lead to an unusable quasi-continuum spectrum, whereas higher inertial defects (up to a few times $10^{-3}$)
lead to the broadening of the ``teeth'' but still conserve the
periodicity of the pattern. We show explicitly in Appendix
\ref{appendix} that for $\epsilon, \delta$ less than these critical
values, the comb ``teeth'' are not only evenly spaced, they fall
precisely (within an error much smaller than a few MHz) at frequencies
\beq
\nu_J = \Delta \nu_{\rm line}\left(J + \frac12\right),
\eeq
where $J$ is an integer and $\Delta \nu_{\rm line} = 2 A_3\left(1 +
  \mathcal{O}(\epsilon^2, \delta)\right)$. 
This very simple pattern will make it possible to use simple
matched-filtering techniques to look for PAH rotational lines (see
Section \ref{sec:comb-fitting}). 

We now turn to estimating the level of asymmetry and inertial defect
of realistic PAHs.

\section{Expected imperfections of realistic PAHs}\label{sec:real-PAH}

\subsection{Characteristic magnitude of the inertial defect}

It is in principle possible to compute the inertial defect $\Delta \equiv I_3
- I_1 - I_2$ of a
symmetric molecule from first principles, see \cite{Jagod_1990} and references therein. The inertial defects of a few small PAHs have been either
measured or computed (see for example \citealt{Thorwirth_2007});
however, for large PAHs such computations become very involved, and to
our knowledge no
experimental nor theoretical values are currently available.

Based on measured inertial defects for a few small aromatic molecules,
\cite{Oka_1995} provides a fit to the inertial defect  in the ground vibrational state, as a function of the
wavelength of the lowest-lying vibrational mode $\lambda_0$ and of the largest moment of inertia $I_3$:
\beq
\frac{\Delta}{1~\textrm{amu \AA}^2} \approx - 0.34
\frac{\lambda_0}{100 ~\mu \rm m} + 0.80 \left(\frac{I_3}{10^4 ~\textrm{amu \AA}^2}\right)^{1/2}.\label{eq:Delta-Oka}
\eeq
The first, negative term comes from the the small non-planarity
induced by out-of plane zero-point vibrations. The second, positive term does
not have a simple classical analog and arises from vibration-rotation
interactions.

Based on the computed rotational constants of coronene and
circumcoronene \citep{Hudgins_2005}, we obtain the following scaling
for the principal moment of inertia for disc-like PAHs:
\beq
I_3 \approx 1.5 \times 10^4 \left(\frac{N_{\rm C}}{54}\right)^2 \textrm{amu
  \AA}^2. \label{eq:I3amu}
\eeq
The lowest-lying vibrational mode of coronene is at $\lambda_0 \approx
80 ~\mu$m, and that of circumcoronene is at $\lambda_0 \approx
180~\mu$m \citep{Bausch_2010, Boersma_2011}. Using Eq.~(\ref{eq:I3amu}), equation
(\ref{eq:Delta-Oka}) gives inertial defects $\Delta \approx 0.2$ and
0.4 amu \AA$^2$ for coronene and circumcoronene, respectively. These values are certainly
highly inaccurate, but one may expect the inertial defect to be of
order a few tenths of amu \AA$^2$, except for molecules with
unusually low vibrational frequencies or unusually low rigidity. For reference, the measured
inertial defect of the symmetrical benzene molecule C$_6$H$_6$ is 0.05 amu \AA$^2$
\citep{Jagod_1990}. On the other hand the following asymmetric
molecules have a negative inertial defect: that of azulene C$_{10}$H$_8$ is -0.15 amu
\AA$^2$, that of acenaphthylene C$_{12}$H$_8$ is -0.19 amu \AA$^2$
\citep{Thorwirth_2007} and that of pyrene C$_{16}$H$_{10}$ is -0.6 amu
\AA$^2$ \citep{Baba_2009}.

Using Eq.~(\ref{eq:I3amu}), the dimensionless inertial defect is of order
\beq
\delta \approx 3 \times 10^{-5} \left(\frac{54}{N_{\rm C}}\right)
\frac{\Delta}{0.4 (N_{\rm C}/54)~ \textrm{amu \AA}^2},
\eeq
where we have normalised the inertial defect to a scaling law fitting
approximately the measured value for benzene and the estimates we
obtained for coronene and circumcoronene.

We conclude that as long as the inertial defect does not exceed
expected values of a few tenths of amu \AA$^2$ for PAHs with $N_{\rm
  C} \sim 50$, its effect on the rotational spectrum is
negligible or at most minor, with $\delta \lesssim 10^{-4}$. Accurate computations of the inertial defect of large PAHs
in order to confirm this statement would be highly valuable.

\subsection{Degree of asymmetry of symmetric PAHs with imperfections}

Let us first consider a simplified model for a compact, symmetric PAH,
which we model as a constant surface-density disc. Assuming a C-C bond
length of 1.4 \AA, Circumcoronene C$_{54}$H$_{18}$ can be modelled as a disc of radius
$a \approx 6$ \AA, having a mass $M\approx 666$ amu, this corresponds to
a surface density $\sigma = M/(\pi a^2) \approx 10^{-7}$ g/cm$^2$, leading to
unperturbed moments of inertia $I_3 = 2 I_1 = 2 I_2 = \frac{\pi}{2} \sigma
a^4 = \frac12 M a^2$.

Let us now assume that we place an impurity of mass $\delta m$ at a
distance $d$ from the centre of mass. The principal axis of lowest
inertia passes through the centre of mass and the impurity (we call
this axis 1), and the second axis is perpendicular to it. The
displacement of the centre of mass leads to a small change in the moment of
inertia quadratic in $\delta m/M$. Neglecting this small term, the moment of inertia
$I_1$ is unchanged with the addition of the mass, whereas $I_2$ is
perturbed by an amount $\delta I_2 = \delta m \times d^2$. The asymmetry
parameter is therefore, to lowest order (using $A_1 \approx A_2
\approx 2 A_3$)
\beq
\epsilon = \frac{A_1 - A_2}{2A_3} = 2 \frac{\delta I_2}{I_3} = 4\frac{d^2}{a^2}
\frac{\delta m}{M}.
\eeq
With $M \approx 12 ~m_p N_C$, where $m_p$ is the proton mass and $N_C$
is the number of carbon atoms, and $N_C \approx 54 (a / 6
\textrm{\AA})^2$, we find
\beq
\epsilon \approx 6 \times 10^{-4} \left(\frac{N_C}{54}\right)^{-2} \frac{\delta m}{2 m_p}
\left(\frac{d}{1.4 \ \textrm{\AA}}\right)^2. \label{eq:epsilon}
\eeq
We have implemented a more realistic PAH model with a honeycomb
skeleton, setting the C-C bond length to 1.4 \AA ~and the C-H bond length to 1.1 \AA, and found that
Eq.~(\ref{eq:epsilon}) gives an accurate estimate of the asymmetry parameter.

We see right away that substituting a peripheral hydrogen by a radical
such as CH$_3$ or adding a benzene ring to the periphery of a
symmetric PAH would lead to an asymmetry of at least several percent and
therefore wash out the comb emission. Only relatively minor changes
($\delta m$ of a few $m_p$) would preserve a sufficient degree of symmetry.  
We now examine what kind of imperfections may lead to an electric
dipole moment, yet preserve a well-identifiable comb spectrum.

\subsubsection{Nitrogen substitution}

\new{Based on the observed position of the 6.2 $\mu$m feature,
\cite{Hudgins_2005} argued that nitrogen-substituted PAHs could represent a significant
fraction of the PAH population, and estimated the N/C substitution
rate in PAHs to be of at least 3\%. Note that this number is uncertain:
for example \cite{Pino_2008} have argued that the shift in the 6.2 $\mu$m feature can also be
explained by the presence of aliphatic bonds. Keeping this caveat in mind, we
shall assume a fiducial N/C substitution rate of 3\% in what follows. Nitrogen-substituted PAHs are
strongly polar, with dipole moments of a few Debyes (see Table 5 of \citealt{Hudgins_2005}), and, according to
Eq.~(\ref{eq:epsilon}), remain symmetric enough to produce sharp combs
of lines. } 

With a substitution rate of 3\%, about 1/3 of coronene and
circumcoronene are singly-substituted. There are 3 and 6 distinct possibilities of N-coronenes and
N-circumcoronenes, respectively (see Fig.~2 of
\citealt{Hudgins_2005} and Table \ref{tab:pahs} of this paper), some with twice the abundance of others due
to a larger phase-space. The number of distinct isomers grows rapidly
with the number of substitutions: there are 29 unequivalent
doubly-substituted coronene ``isomers'', and over 100 kinds of doubly-substituted
circumcoronene. Even if these species tend to have a larger dipole
moment (though it may also vanish if the two nitrogens are substituted at symmetric locations), they tend, firstly, to
have a larger asymmetry parameter. Secondly, and more importantly, the large
number of different ``isomers'' dilutes the power that is radiated in
any single comb. The rotational emission of multiply-substituted PAHs
therefore add up to a quasi-continuum, whereas singly-substituted
molecules produce isolated combs. The latter represent our best
candidate for a detection. We show the dipole moments computed by \cite{Hudgins_2005} and our estimated
asymmetry parameters for singly-substituted coronene
and circumcoronene in Table \ref{tab:mu-eps}.

\begin{table}
\caption{Properties of singly nitrogen substituted coronene and
   circumcoronene. The numbering of substitution positions corresponds
   to that shown in Table \ref{tab:pahs}. The dipole moment is obtained
   from \citealt{Hudgins_2005}. The asymmetry parameter is approximate and
   estimated assuming a fixed honeycomb carbon skeleton with 1.4
 \AA~C-C bonds and 1.1 \AA~C-H bonds.
The fractional abundance is with respect to the total amount of singly-substituted coronene (or circumcoronene) made entirely of C$^{12}$ and in a given charge state.
} \label{tab:mu-eps}

\begin{tabular}{cccc} \hline\hline 
 &N-Coronene& (NC$_{23}$H$_{12}$)&  \\
\hline\hline
Position &$\mu$ (Debye) & Asymmetry $\epsilon$ & Fraction \\
\hline 
1&2.67 & 0.003 & 1/4 \\
2&3.69 & 0.010 & 1/4 \\
3&5.48 & 0.018 & 1/2 \\
\hline\hline 
& N-Circumcoronene &  (NC$_{53}$H$_{18}$)&\\
\hline\hline
Position &$\mu$ (Debye) & Asymmetry $\epsilon$ & Fraction \\
\hline 
1&1.32 & 0.0005 &1/9 \\
2&4.55 & 0.0021 &1/9 \\
3&5.43 & 0.0036 &2/9 \\
4&6.79 & 0.0067 &1/9 \\ 
5&6.99 & 0.0083 &2/9 \\
6&9.23 & 0.0100 &2/9 \\
\hline \hline
\end{tabular}
\end{table}

\subsubsection{Dehydrogenation / super-hydrogenation}

Under the harsh ISM conditions, small PAHs may loose several, and up to
all of their peripheral hydrogen atoms. \cite{LePage_2003} found that
PAHs with $\sim 20-30$ carbon atoms are typically stripped of most of
the peripheral hydrogens (often all of them), whereas larger PAHs
typically have normal hydrogen coverage, the transition being a sharp
function of size. As long as peripheral
hydrogen atoms are either completely stripped off or all in place, the
resulting molecule remains symmetric and non-polar. If one peripheral
hydrogen is missing, or if there is only one H present, of if there is
one additional H on top of a full coverage, the PAH
acquires a permanent dipole moment as well as a small asymmetry. The
loss or addition of more than one hydrogen atom leads to a large
number of sub-families among which the power in rotational lines is
spread. Given the sharp transition from full-hydrogenation to
full-dehydrogenation, it is likely that the species that could be of interest,
namely singly (de- or super-) hydrogenated PAHs represent a small
fraction only of the total PAH abundance. We shall make the optimistic
assumption that nitrogen-substituted PAHs are mostly either fully
hydrogenated or completely dehydrogenated.

\subsubsection{\new{$^{13}$C} substitution}

The interstellar $^{12}$C/$^{13}$C ratio is approximately 70 (see
\citealt{Ritchey_2011} and references therein). Assuming the isotopic ratio is the
same among PAH carbon atoms, this implies that $\sim 30\%$ of
coronene and $\sim 55\%$ of circumcoronene contain one $^{13}$C atom or more. $^{13}$C-substitution by itself does
not lead to any significant dipole moment: if the substitution is at a
distance $d$ from the centre of a symmetric PAH of mass $M$, the centre of mass is
displaced by $\sim (m_p/M) d$ with respect to the centre of charge,
which implies a dipole moment of a few $10^{-3}$ Debyes at most for a
PAH of $\sim 50$ carbon atoms. However, $^{13}$C substitution does
change the rotational constant of the bearing PAH. Polar nitrogen-substituted
PAHs which in addition have one or more $^{13}$C in their skeleton
therefore get subdivided in a large number of sub-families, depending on the
relative position of the N and $^{13}$C atoms in the skeleton, each
with different rotational constants and spectrum. They therefore contribute to the quasi-continuum spinning
dust emission. Only nitrogen-substituted PAHs that are made purely of $^{12}$C are
susceptible of having strong combs, that is, $\sim 70\%$
of coronene and $\sim 45\%$ of circumcoronene.

\subsubsection{Deuterium substitution}

The measured D/H ratio varies significantly from one sightline to
another, from about $\sim 7$ ppm to values closer the primordial ratio
of $\sim 26$ ppm \citep{Linsky_2006}. Astration alone cannot easily
explain such variations, due to the lack of correlation with variation
in the O/H ratio, and \cite{Draine_2006} suggested that deuterated
PAHs may contain a significant fraction of the interstellar
deuterium. Measurements the weak 4.65 $\mu$m feature characteristic of
the C-D stretching mode in deuterated PAHs indeed indicate that the D/H ratio in
PAHs could be as high as $\sim 0.3$ \citep{Peeters_2004}. If this is
the case, all PAHs will have at least a few deuterium atoms in their
periphery, resulting in a large number of families with different
rotational constants, and diluting all combs into a
quasi-continuum. This hypothesis remains to be firmly confirmed,
however, and in any case not all sight-lines show a low D/H ratio, so
one can expect that in some cases PAHs are mostly
deuterium-free. We will make this assumption but keep in mind that
deuterium-substitutions may very well significantly dilute our
predicted signal.

\subsubsection{Charge states}

Different charge states should have slightly different rotational
constants. Even a relative difference of $10^{-4}$ would lead to a
shift of lines by a few MHz at a few tens of GHz. Each N-PAH is
therefore divided among several charge states with a priori different
rotational constants. Luckily, small PAHs
are found mostly in two states (see for example Fig.~1 of
\citealt{DL98_long}), i.e. this implies a reduction of each comb
strength by a factor of $\sim 2$ only.

\subsection{Conclusion of this section}\label{sec:small-conclusion}

Let us summarise the essential conclusions of this section:

$\bullet$ The spectrum of an ideal planar and highly symmetrical PAHs with D$_6$h or D$_3$h
symmetry would appear as a perfect ``comb'' of line stacks, each one
being made of a large number of lines sharing the same frequency.

$\bullet$ A small asymmetry or inertial defect breaks the
frequency degeneracy and spread the radiated power of each stack
over a finite frequency interval. Provided these defects are small
enough, each stack may still appear as a strong ``line'' when observed
with a few MHz resolution. Inertial defects no larger than several tenths of
amu \AA$^2$ and asymmetries of a percent or less are small enough to preserve the comb structure of the observed spectrum.

$\bullet$ The majority of any PAH ``family'' (such as coronene and all
substituted versions of it) is divided into a very large number of
different ``isomers'' with different rotational constants. These
isomers may be quite symmetric and each emit a comb of lines, but in
general the abundance in
any one of them is small so their combined rotational emission makes
up a quasi-continuum unusable for the purposes of identification.

$\bullet$ A few special isomers are at the same time highly polar,
highly symmetric, and over-abundant: these are the singly
nitrogen-substituted PAHs, made exclusively of C$^{12}$, and not
suffering from any other defect such as deuterium substitution or
dehydrogenation. Assuming the latter two defects are rare (and keeping
in mind that their actual rate of occurrence is very uncertain), the
largest fraction of coronene in a single ``isomer'' is $\sim 0.7
\times 0.5 \times 0.36 \times \frac12 \approx 6 \%$, where the
factors are the fraction of pure-C$^{12}$ coronene, the fraction in
each of the two main charge states, the fraction of
single nitrogen-substitution, and the fraction of those in the most abundant
isomer. These fractions decrease with size and the corresponding number for circumcoronene is $\sim
0.45 \times 0.5 \times 0.32 \times \frac29 \approx 1.6 \%$. For PAHs
with D$_3$h symmetry (the approximately triangle-shaped PAHs in Table
\ref{tab:pahs}) instead of D$_6$h symmetry (the approximately
hexagon-shaped PAHs, such as coronene and circumcoronene), the number
of ``isomers'' is about twice as large, and this fraction is twice as small.

\section{Searching for combs in the spinning dust forest} \label{sec:comb-fitting}

The idea to search for the rotational emission of specific symmetric
top PAHs has already been suggested by \cite{Lovas_2005}, and a search
for corannulene C$_{20}$H$_{10}$ has been carried by \cite{Pilleri_2009} in the Red
Rectangle, who set an upper limit $N_{\rm C_{20} \rm H_{10}}/N_{\rm H}
< 3 \times 10^{-11}$ by targeting the specific $J =
135\rightarrow 134$ transition at 137.615 GHz. 

Looking for a single line, however, is suboptimal, as the peculiarity
of rotation spectra of a planar symmetric top, is that the rotational lines fall
almost exactly at positions $\nu = (J + 1/2) \Delta \nu_{\rm
  line}$. The purpose of the last section was to demonstrate that for
small enough asymmetry and inertial defect, this pattern is preserved,
and that singly nitrogen-substituted PAHs do satisfy these requirements.

Rather than searching for a specific line, a better search strategy is to use the
largest possible bandwidth with the best possible resolution, and
search for \emph{comb patterns}, resulting in a much enhanced signal-to-noise ratio (by a factor of the
square root of the number of lines observable within the total bandwidth). This has
the additional advantage of allowing for a blind search over a large
range of values of $\Delta \nu_{\rm line}$, hence over a broad range
of carrying species, rather than focusing on a single species at a
time.

In what follows we illustrate a simple method to extract a comb signal
from a noisy spectrum, identical in spirit to mathed-filtering methods
used for the search of gravitational wave signals. We consider $N_{\rm bins}$ flux measurements $d_i$ at frequencies $\nu_i$, with constant
resolution $\Delta \nu_{\rm res}$, and over a total bandwidth $\Delta
\nu_{\rm tot} = N_{\rm bins} \Delta \nu_{\rm res}$. We
assume that we have subtracted the broadband average from the signal,
obtained for example by averaging the data over 1 GHz bands. The
measured signal is noisy, with a noise $n_i$, which we assume to be uncorrelated with a known variance: $\overline{n_i n_j} = \mathcal{N}^2
\delta_{ij}$. The derivation below can be easily generalised to
frequency-dependent and correlated noise using the noise covariance
matrix.

\begin{figure}
\includegraphics[width = 85 mm]{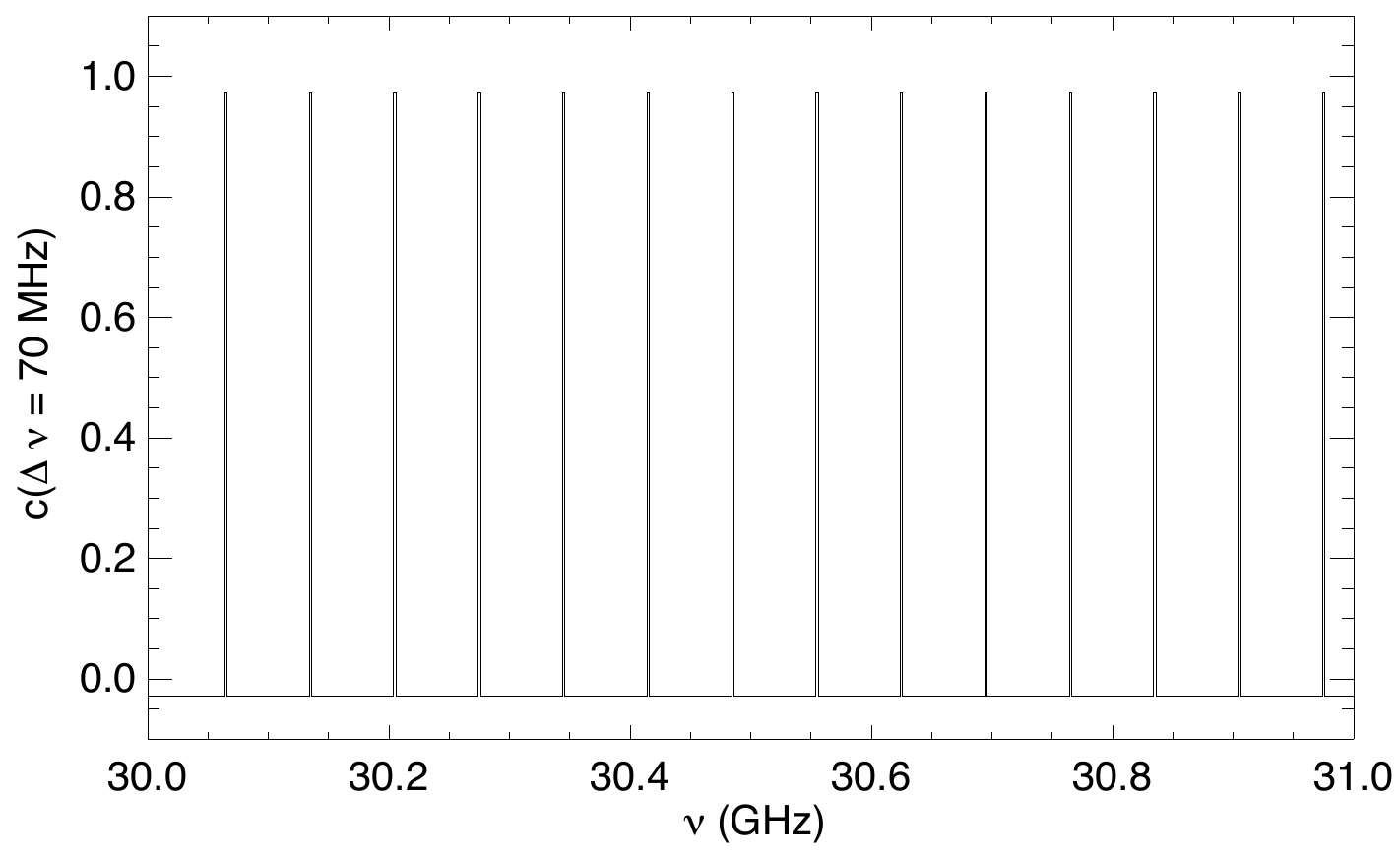}
\caption{Dimensionless comb basis function $c(\Delta \nu = 70$ MHz), for frequency
  bins $\Delta \nu_{\rm res} = 2$ MHz. The base is offset from zero because we have
  subtracted the mean.}\label{fig:basis} 
\end{figure}

Let us assume that the signal contains $N_{\rm combs}$ combs (labelled
with subscripts $a, b$) with respective line spacings
$\Delta \nu_{\rm line}(a) = \Delta \nu_a$, and peak flux density $s^a_i$ (where we have again subtracted the
broadband average from the flux in the combs), so the data can be written as
\beq
d_i = n_i + \sum_a s^a_i. \label{eq:di}
\eeq
For two functions $f, g$ with zero mean, we define the scalar product 
\beq
\langle  f \cdot g \rangle \equiv \frac1{N_{\rm
    bins}}\sum_{i=1}^{N_{\rm bins}} f_i g_i, \label{eq:dot.prod}
\eeq
which is just the average of their product over the observed
bandwidth, and the associated norm $|| f|| \equiv \langle f^2
\rangle^{1/2}$. Providing the bandwidth is no more than a few GHz, the
amplitude associated with each comb should be approximately constant across
the observed frequency range. The underlying signal can therefore be
decomposed onto the infinite set of dimensionless ``comb'' basis
functions $c_i^a \equiv c_i(\Delta \nu_a)$, which are unity in
frequency bins containing a line and zero otherwise, minus the
resulting mean (i.e. $c_i^a \approx 1 - \frac{\Delta \nu_{\rm
    res}}{\Delta \nu_a}$ if there exists an integer $J$ such that $\nu_i - \Delta
\nu_{\rm res}/2 < (J + 1/2) \Delta \nu_a <  \nu_i + \Delta
\nu_{\rm res}/2$ and $- \frac{\Delta \nu_{\rm
    res}}{\Delta \nu_a}$ otherwise). We show an
example in Fig.~\ref{fig:basis} for clarity. The signal can therefore
be approximately rewritten as 
\beq
s_i^a \approx \mathcal{A}_a c_i^a, 
\eeq
where $\mathcal{A}_a$ is the peak flux density.
The comb basis functions are nearly orthogonal for the scalar product
(\ref{eq:dot.prod}): if two spacings $\Delta \nu_a$ and $\Delta \nu_b$
are not harmonics of one another, the scalar product $\langle c^a
\cdot c^b \rangle $ is typically much smaller than $||c^a|| \times ||c^b||$. It is not in general
exactly zero because of the finite bandwidth, but is \emph{on average}
zero (when keeping $\Delta \nu_a$ fixed and varying $\Delta \nu_b$
for example), and has a variance
\beq
\overline{\langle c^a \cdot c^b\rangle ^2}  = \frac{||c^a||^2
  ||c^b||^2}{N_{\rm bins}} \ll ||c^a||^2
  ||c^b||^2, \ \ \textrm{if} \ a \neq b. 
\eeq
The norm of each comb basis function is
\beq
||c^b||^2 \approx \frac{N_{\rm lines}^b}{N_{\rm bins}} = \frac{\Delta
  \nu_{\rm res}}{\Delta \nu_b},
\eeq
where $N_{\rm lines}^b \equiv \frac{\Delta \nu_{\rm tot}}{\Delta \nu_b}$
is the number of lines of the comb $b$ within the full bandwidth. 

The matched-filtering consists in computing the following quantity as
a function of comb spacing:
\beq
\mathcal{S}(\Delta \nu_b) \equiv \frac{\langle d \cdot c^b \rangle
}{||c^b||^2} = \frac{\langle n \cdot
  c^b\rangle }{||c^b||^2} + \sum_a \mathcal{A}_a \frac{\langle c^a \cdot
  c^b \rangle}{||c^b||^2}.
\eeq
The mean of $\mathcal{S}$ is the amplitude of the comb signal with
spacing $\Delta \nu_b$, providing such a comb exists in the
data\footnote{We have ignored the fact that $\mathcal{S}(\Delta \nu_b)$ is also
non-zero if the signal contains harmonics of $\Delta \nu_b$. We have found that the dominant
harmonics are combs with 3, 1/3, 5 and 1/5 times the fundamental
spacing, with contributions a factor of a few smaller than that of the
principal value. One could account for this fact (and utilise it to extract more information) with a more sophisticated
data analysis technique, but this is beyond the scope of the present
work, and we shall neglect resonances for simplicity.},
\beq
\langle \mathcal{S} (\Delta \nu_b)\rangle  = \mathcal{A}_b,
\eeq
and its variance in the absence of signal at $\Delta \nu_b$ is is the sum of the instrumental noise and the
contribution from the signal itself:
\beq
\langle \mathcal{\delta S}^2 \rangle =\frac1{N_{\rm bins} ||c^b||^2} \left[\mathcal{N}^2 +
  \sum_{a} ||s^a||^2\right] = \frac{||d||^2}{N_{\rm lines}^b},\label{eq:noise}
\eeq
where $||d||$ is just the rms fluctuation of the total data (signal
\emph{and} noise). The signal-to-noise ratio (hereafter, SNR) is therefore
\beq
\textrm{SNR} (\Delta \nu_b) = \frac{\langle \mathcal{S} (\Delta \nu_b)\rangle}{\langle \delta
  \mathcal{S}^2 \rangle^{1/2}} = \frac{\sqrt{N_{\rm lines}^b}
  \mathcal{A}_b}{\langle d^2 \rangle^{1/2}},
\eeq
i.e. it is enhanced by a factor of $\sqrt{N_{\rm lines}^b}$ with respect to the
single-line signal-to-noise ratio. For a bandwidth of order 10 GHz and
a line spacing of order 100 MHz, this enhancement is a factor of 10.

Since the noise per bin is $\mathcal{N} \propto 1/\sqrt{\Delta
  \nu_{\rm res}}$ (and similarly for $||d||$) and the flux density in each line is
  inversely proportional to the bin width, $\mathcal{A} \propto
  1/\Delta \nu_{\rm res}$, we see that the signal-to
  noise ratio increases proportionally to $\sqrt{\Delta \nu_{\rm tot}/\Delta \nu_{\rm
      res}}$. Even if the noise level increase when increasing the
  resolution, the line intensity increases faster. It is therefore
  best to have as large a bandwidth as possible and as high a resolution
  as possible. In practice, however, there is no point in having a
  resolution better than the turbulent velocity broadening, of
  magnitude $\Delta \nu_{\rm turb} =$ 1 MHz$\times(v_{\rm turb}/10$ km/s) at 30 GHz.

\begin{figure*}
\includegraphics[width = 85 mm]{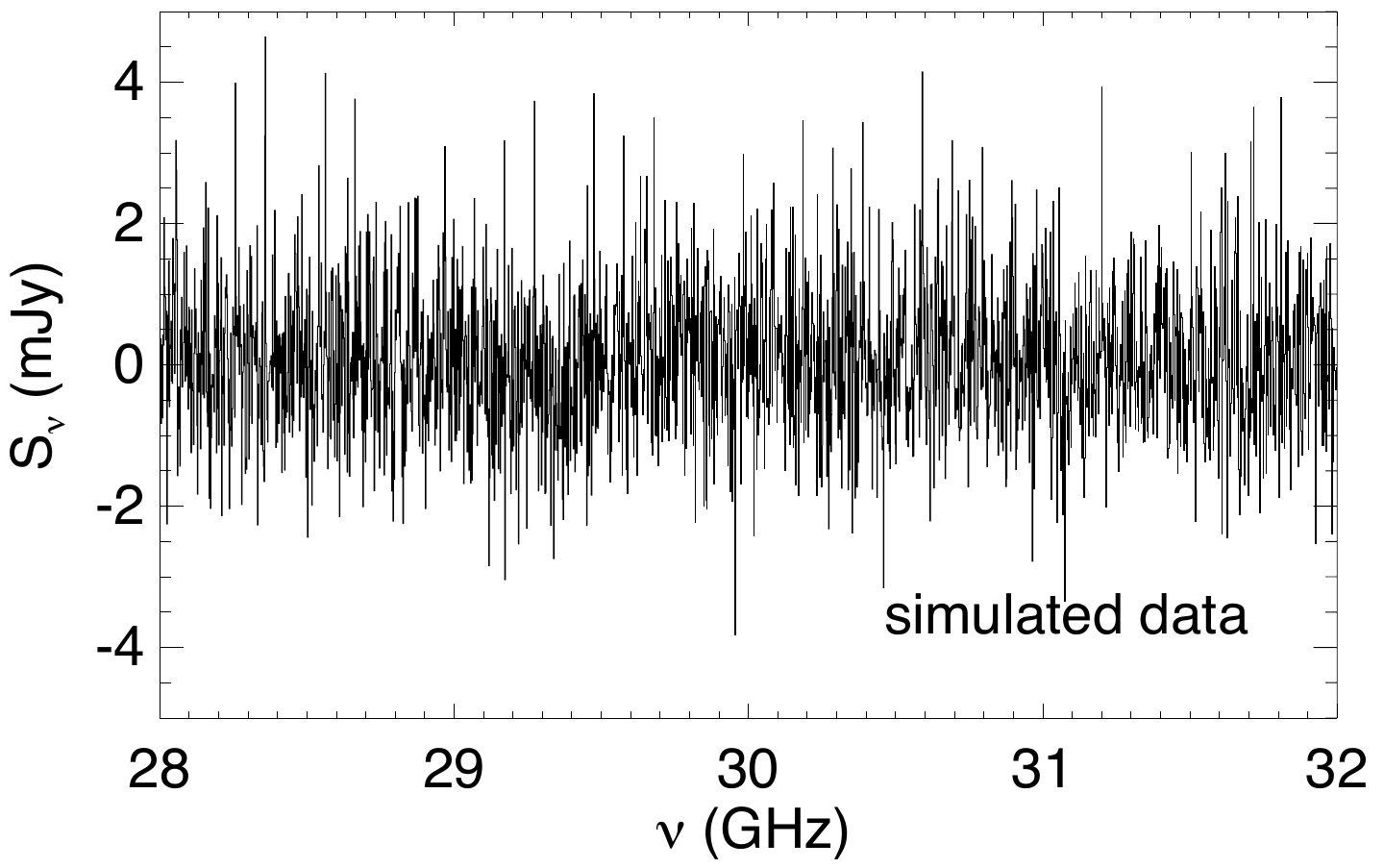}
\includegraphics[width = 85 mm]{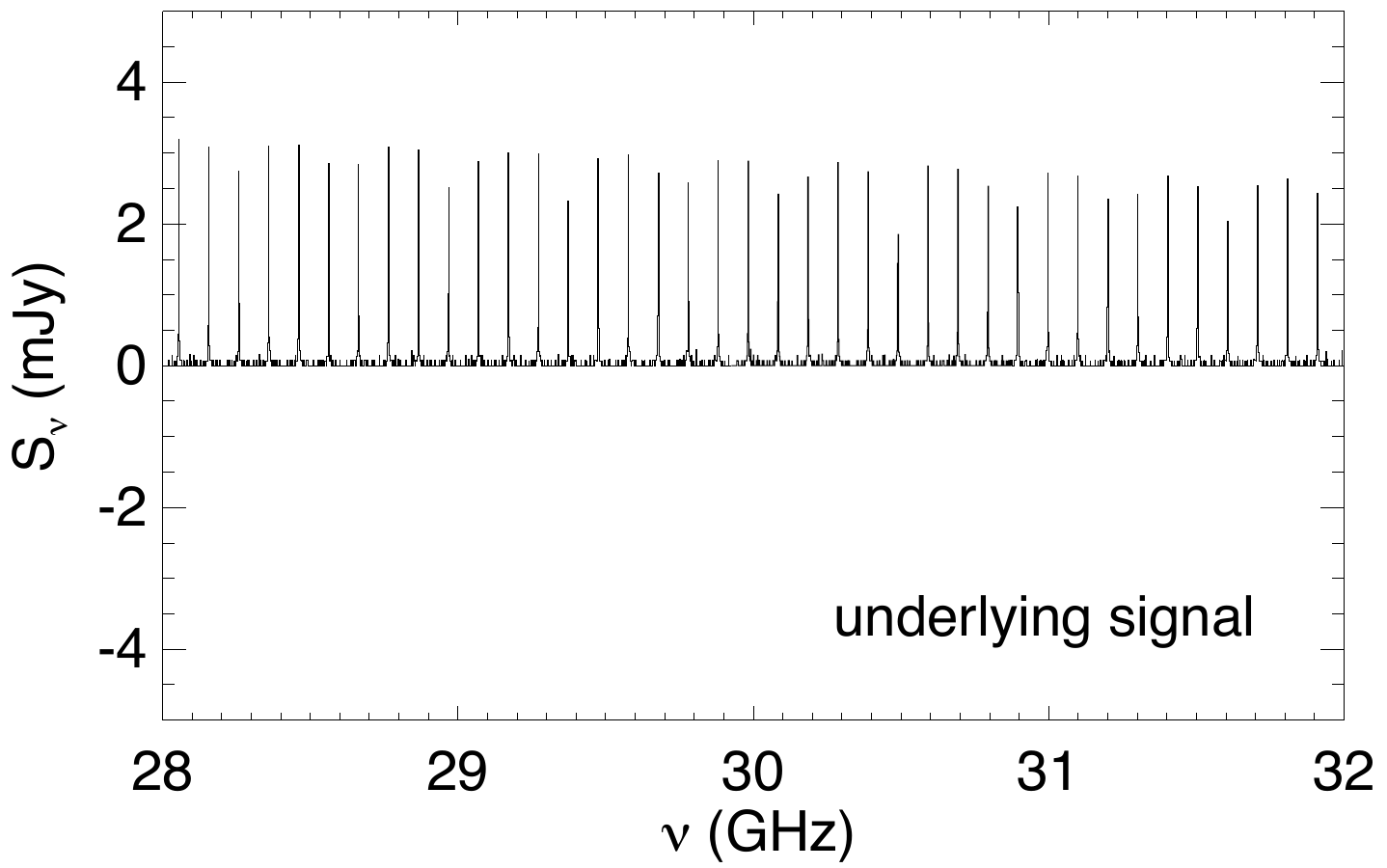}
\caption{Simulated data, normalised to the rms noise
  value, assumed uniform across the full bandwidth. Observations are
  over a 10 GHz total bandwidth (only 4 GHz are shown for better clarity), with a 2 MHz
  resolution. The assumed underlying signal is shown in the right panel, and was computed assuming a rotational constant $2 A_3 = 101.46$ MHz (corresponding to $N_{\rm C} \sim 40$), an asymmetry $\epsilon = 3 \times 10^{-3}$, no
  inertial defect, a dipole moment $\mu = 2$ Debye, and a Maxwellian
  distribution for $J$ such that the emission peaks around 30 GHz.}\label{fig:data-sim} 
\includegraphics[width = 83 mm]{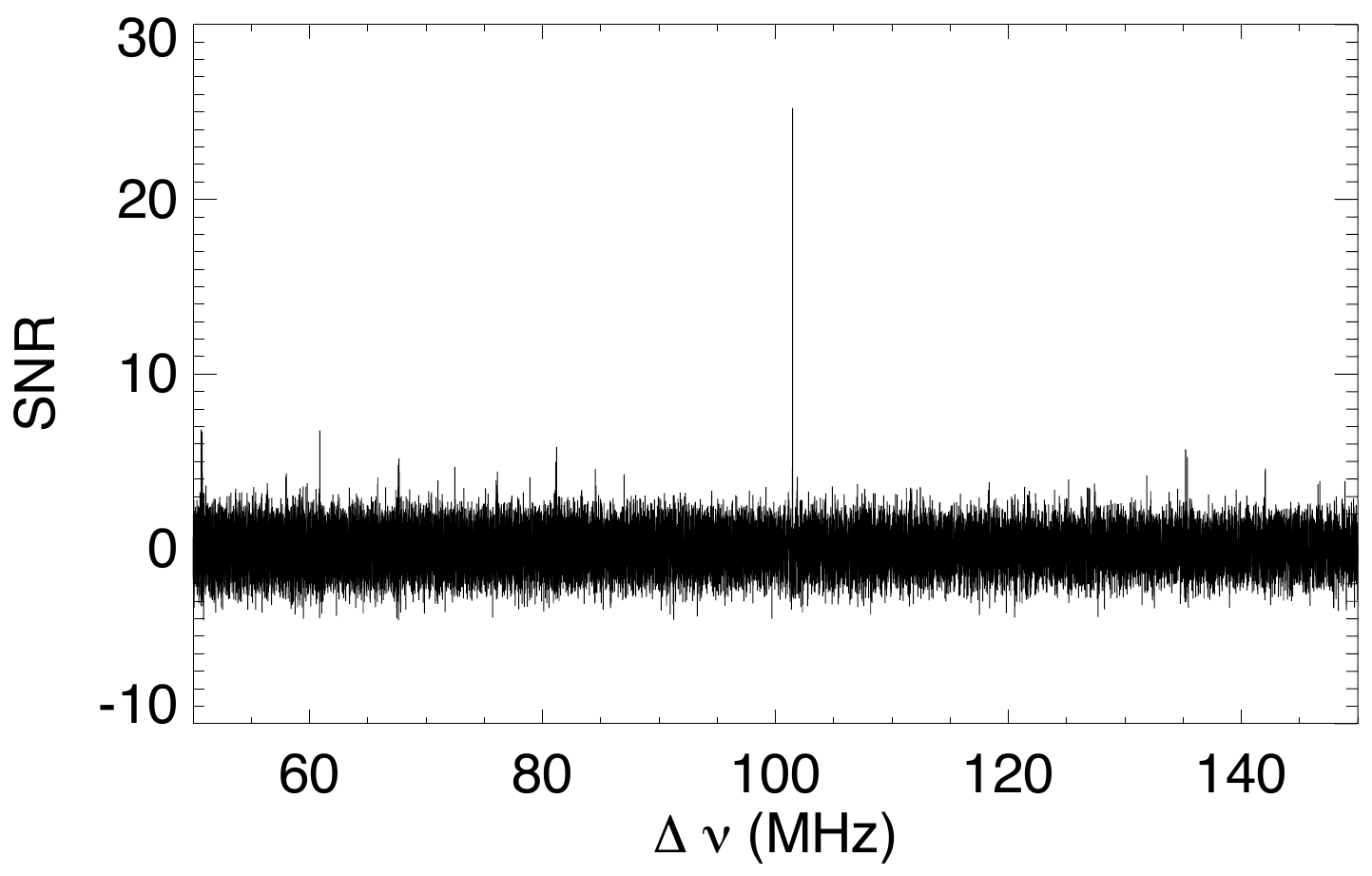}
\includegraphics[width = 87 mm]{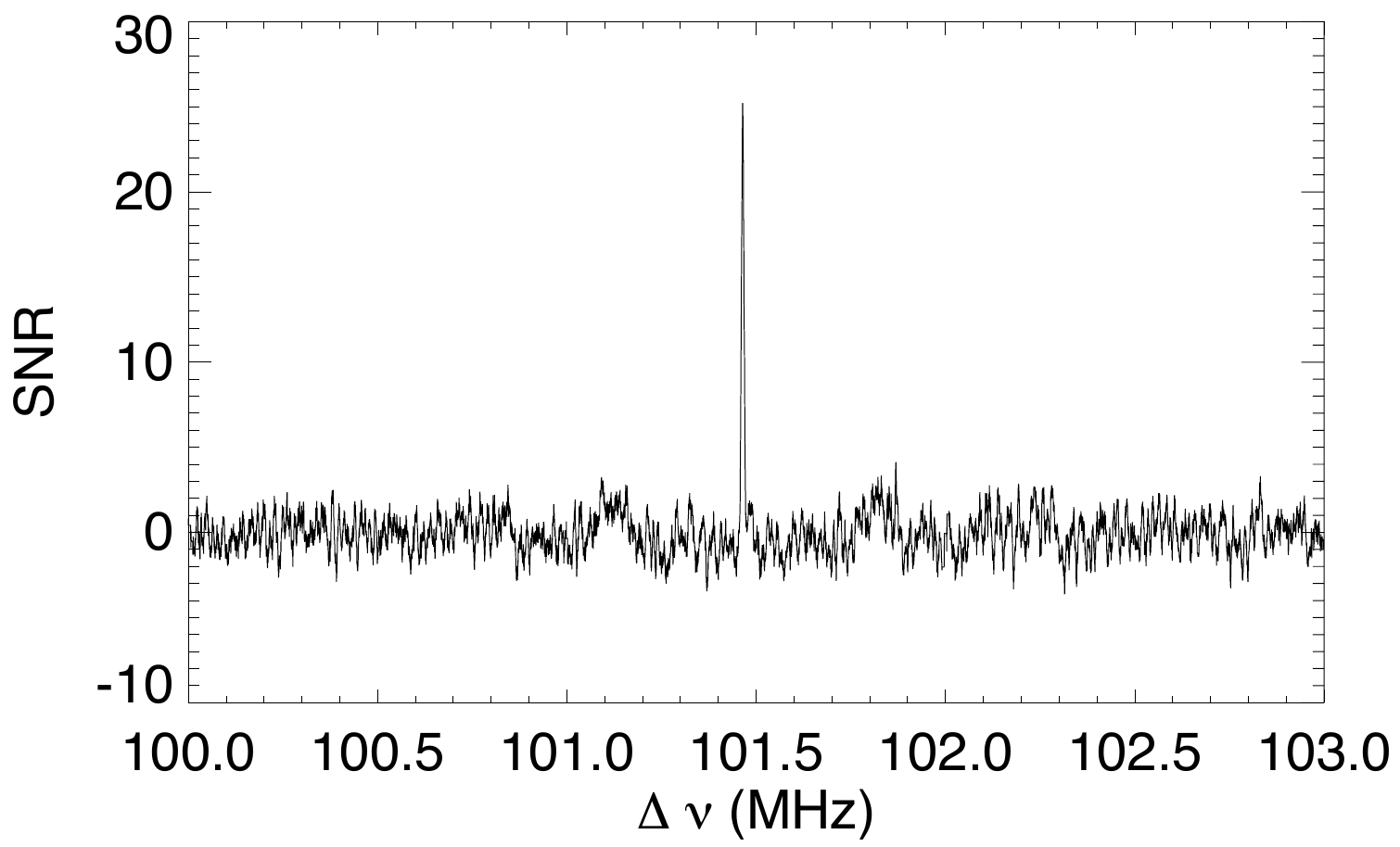}
\caption{Signal-to-noise ratio obtained with the matched-filtering
  technique described in this Section, as a function of trial line
  spacing $\Delta \nu$, and using the data shown in
  Fig.~\ref{fig:data-sim}. The right panel is a zoom near the $\Delta
  \nu \approx$ 100 MHz region. The matched-filtering
  technique allows to detect the underlying comb with SNR of 25, and
  to recover the input frequency with very high accuracy.}\label{fig:SNR} 
\end{figure*}

We illustrate the method in Figs.~\ref{fig:data-sim} and
\ref{fig:SNR}. We have simulated the rotational spectrum of a $\sim
40$-carbon atom planar PAH, with rotational constant $A_3
\approx 50$ MHz, asymmetry parameter $\epsilon = 3\times 10^{-3}$,
no inertial defect, and permanent dipole moment $\mu = 2$ Debye. We
assumed a probability distribution $P(J, K) \propto J
\exp[-(J/90)^2]$, so the spectrum peaks near 30 GHz. We convolved the
spectrum with a 2 MHz window function over a 10 GHz bandwidth (Fig.~\ref{fig:data-sim}, right panel), and added a
gaussian random noise with a rms value of slightly more than a third of the mean line
amplitude (Fig.~\ref{fig:data-sim}, left panel). With this
configuration, no individual line could have been detected at the
3-$\sigma$ level. We performed a blind
search over comb spacings $\Delta \nu$, as described above. The SNR as
a function of $\Delta \nu$ is shown in Fig.~\ref{fig:SNR}. We see
that in this case the input comb is detected with a very high SNR (about 25 in this
example), and the line spacing is recovered with very high accuracy.

Let us finally point out that there is an intrinsic ``noise'' to the signal
originating in its non-smoothness, represented by the $\sum_a ||s^a||^2$ term in Eq.~(\ref{eq:noise}). If there are too many combs with
comparable amplitudes, the signal may indeed disappear under the
grass. For $N_{\rm comb}$ combs with similar amplitude, the ``noise''
due to the lines themselves is $\left(\sum_a ||s^a||^2\right)^{1/2} \approx
  \sqrt{N_{\rm comb}} \mathcal{A} \sqrt{N_{\rm lines}^b/N_{\rm bins}}$, which would become
  comparable to $\sqrt{N_{\rm lines}^b} \mathcal{A}$ if $N_{\rm comb} \sim
  N_{\rm bins}$, i.e., for a bandwidth of several GHz and a few MHz resolution, if there
  are more than $\sim 1000$ species with comparable abundance emitting
  sharp comb signals. While this seems unlikely, should this be the
  case, if the intrinsic ``noise'' due to the PAH combs is
  significantly larger than the instrumental noise, one should still
  be able to detect the presence of small PAHs from the anomalous
  noise level. This would require an accurate calibration of
  the instrumental and atmospheric noise by switching regularly to a relatively empty field of view with
no expected PAH emission. Whereas this is in principle possible, the
practical implementation of such a search is likely to be challenging,
and so would be the interpretation of the results. We have therefore not analysed
this option in more detail in this paper.

\section{Detection forecasts}\label{sec:forecast}

\subsection{Expected SNR from the observed spinning dust emission}

One can in principle predict the intensity of the rotational lines
of any given PAH given the local environment (density, temperature,
ambient radiation field, etc...). The first step is to compute the distribution $P(J, K)$ for the
angular momentum and configuration of the grain (see for example
\citealt{spdust_review} for a review of the theory). The spectrum is
then readily obtained given the asymmetry parameter and inertial defect, using
Eqs.~(\ref{eq:J-1})-(\ref{eq:J-1Kpm1}) for the power radiated in each
line, for which the exact frequency is easily computed from the
expression for the energy levels (\ref{eq:EJK-final}). Code to do
these computations using output from \textsc{SpDust} \citep{spdust1} can be obtained
from the author.

It is however difficult to estimate the exact conditions along a given
line of sight (and these conditions are moreover non-uniform), and we have chosen instead in this work to estimate the
rotational line intensity from the observed anomalous microwave
emission (hereafter, AME). The latter is indeed believed to be
due to spinning dust radiation which is the broadband equivalent of
the rotational line emission we are searching for. 

The first step is to relate the line intensity of any given specific
PAH ``isomer'' (such as neutral NC$_{23}$H$_{12}$ with nitrogen in the
innermost ring) to the overall broadband emission from all its
``family'' (continuing with our previous example, all the substituted versions of coronene, including
multiple nitrogen substitutions, isotope substitutions, and different
charge states). The intensity of combs
produced by each isomer decreases with increasing asymmetry and depends on the
dipole moment $\mu$ in two ways: first, the power radiated is proportional
to $\mu^2$, second, the characteristic rotation rate hence the peak
of the emission decreases with increasing dipole moment due to the radiation reaction
torque, see for example Fig.~12 of \cite{spdust1}. The dipole moment
increases with the asymmetry parameter (see Table \ref{tab:mu-eps}),
and theses two quantities have approximately an opposite effect on the line
intensity (provided the observation frequency is near the peak of the
emission). Guided by this observation, we shall assume that the line
intensity is approximately constant among different isomers, and
equal to that of the most symmetric isomer. We have currently no observational handle on the
characteristic dipole moment of PAHs. \cite{DL98_long} assume that $\mu \approx 0.4$ Debye $\times \sqrt{N}_{\rm
  atoms}$, based on the known dipole moments of a few
small carbonaceous and nitrogenated radicals. The reasonable agreement
of the \cite{DL98_long} spinning dust model with observations
indicates that this characteristic dipole moment gives, at least within a
factor of a few, a relatively accurate representation of real polar
PAHs. This simple scaling gives $\mu \approx 2.4$ Debye for coronene
and 3.4 Debye for circumcoronene, which is comparable to the dipole
moment of their most symmetrical nitrogen-substituted isomers, where
the nitrogen atom is in one of the innermost rings. From these considerations, we shall assume that intensity in the lines
of any isomer is comparable to the intensity of a \new{PAH} with no
asymmetry and with a ``typical'' dipole moment, i.e. comparable to the
rms dipole moment of the PAHs producing the broadband spinning dust
emission. We may therefore relate the flux in any line due to a
particular isomer to the broadband flux emitted by its entire
``family'' as
\beq
S_{\nu}^{\rm line}(\textrm{isomer}) \sim \frac{N_{\rm isomer}}{N_{\rm family}}\frac{\Delta \nu_{\rm line}}{\Delta
  \nu_{\rm res}} S_{\nu}^{\rm bb}(\textrm{family}),
\eeq
where $\Delta \nu_{\rm line}$ is the line spacing for the PAH
considered and $S_{\nu}^{\rm bb}$ is the broadband emission resulting
from the whole ``family'' of similar isomers. 

The next step is to relate the broadband emission from a single PAH
``family'' to the total spinning dust emission, $S_{\nu}^{\rm AME}$,
which is sourced by all (or at least the smallest) of the PAHs. We denote by $N_{\rm PAH, tot}$ the
total abundance of PAHs responsible for the AME -- more precisely, the abundance of PAHs with
$\sim 50$ carbon atoms that would contain the same amount of carbon as the
whole radiating PAH population. The range of sizes spanned by
the PAHs carrying the AME is relatively limited (see for example
Fig.~11 of \citealt{spdust1}), and we may assume that 
\beq
S_{\nu}^{\rm bb}(\textrm{family}) \sim \frac{N_{\rm family}}{N_{\rm
  PAH, tot}} S_{\nu}^{\rm AME}.
\eeq

We therefore arrive at the following simple relation between the intensity
in the lines of any specific substituted PAH and the broadband AME:
\barr
&&S_{\nu}^{\rm line} \sim \frac{\Delta \nu_{\rm line}}{\Delta
  \nu_{\rm res}}  f_{\rm iso} \frac{N_{\rm family}}{N_{\rm
  PAH, tot}} S_{\nu}^{\rm AME} \nonumber\\
&&\hspace{-0.5cm}\sim 0.1 ~\textrm{mJy} \frac{\Delta \nu_{\rm line}}{100 ~\rm MHz} \frac{1~
\rm MHz}{\Delta   \nu_{\rm res}}\frac{f_{\rm iso}}{1\%}
\frac{N_{\rm family}}{0.01~N_{\rm
  PAH, tot}} \frac{S_{\nu}^{\rm AME}}{10~ \rm mJy}, \label{eq:line-from-ame}
\earr
where $f_{\rm iso} \equiv N_{\rm isomer}/N_{\rm family}$ is the
fraction of PAHs of a given family in a specific isomer, and is
typically of order a few percent (see discussion in Section \ref{sec:small-conclusion}).

The resulting SNR with our matched-filtering analysis and with noise
$\mathcal{N}$ per resolution element would then be
\barr
\textrm{SNR} &\approx& \sqrt{\frac{\Delta \nu_{\rm tot}}{\Delta \nu_{\rm line}}}
  \frac{S_{\nu}^{\rm line}}{\mathcal{N}}\\
&\sim& 10 \sqrt{\frac{\Delta
      \nu_{\rm tot}}{10 ~\textrm{GHz}}}\frac{1~ \textrm{MHz}}{\Delta \nu_{\rm
  res}} \frac{0.1 ~\textrm{mJy}}{\mathcal{N}} \nonumber\\
 &\times& \sqrt{\frac{\Delta \nu_{\rm line}}{100~ \textrm{MHz}}} \frac{f_{\rm sym}}{1\%}\frac{N_{\rm PAH}}{0.01~N_{\rm PAH,
  tot}}\frac{S_{\nu}^{\rm
AME}}{10~ \textrm{mJy}}.  \label{eq:forecast}
\earr
One should keep in mind that Equation (\ref{eq:forecast}) is simply
indicative and probably only accurate within an order of magnitude.

For reference, the Spectrometer backend on the Green Bank Telescope
(GBT) can allow for a 3.2 GHz bandwidth, with a sub-MHz
resolution (but again, we only require $\Delta \nu_{\rm res}\gtrsim$ 1
MHz as beyond this resolution Doppler broadening is likely to smear
the lines anyway). The noise level for the KFPA receiver near 26 GHz is approximately\footnote{https://dss.gb.nrao.edu/calculator-ui/war/Calculator\_ui.html}
\beq
\mathcal{N}^{\rm GBT} \approx 0.7~ \textrm{mJy} \sqrt{\frac{1~ \textrm{MHz}}{\Delta
    \nu_{\rm res}} \frac{1 ~\textrm{hour}}{\tau_{\rm obs}}}, \label{eq:GBT-noise}
\eeq
and a noise level of $\sim 0.1$ mJy per 1 MHz resolution element can in principle be achieved after
$\sim 50$ hours of observations. \new{The Very Large
Array\footnote{http://www.vla.nrao.edu/} (VLA) also allows for a sub-MHz
frequency resolution over a large bandwidth (8 GHz in the K band),
with a noise level better than (but comparable to) that of the
GBT. However, because the VLA is an interferometer, it is not
sensitive to power on angular scales larger than $\lambda/D_s$, where
$\lambda$ is the observation wavelength and $D_s$ is
the shortest separation in the array configuration. In the VLA C or D
configurations, the largest angular scale accessible is of order 1
arcmin\footnote{https://science.nrao.edu/facilities/vla/docs/}. It
is therefore more appropriate to use the GBT for extended sources
whereas the VLA would be more suitable for sources that are known to
be compact. Finally, we note that quasi-periodic features in the
instrument noise could deteriorate the sensitivity to ``combs''
compared to that predicted in this idealized analysis. Any such
features should be accurately characterized prior to a search for PAH lines.}

\subsection{Potential targets}

Anomalous microwave emission has been detected along several lines of
sight, and in very different regions of the ISM. A systematic search
for lines should be conducted in known AME regions, 
as well as in regions showing strong PAH infrared emission. 
Here we simply point out a couple of particularly
bright AME sources:

Observations at 31~GHz with CBI reveal the presence of
strong AME in the HII region \textbf{RCW175} ~\citep{Dickinson_2009,
  Tibbs_2012}, with a measured flux of 1 Jy per 4' beam. This
corresponds to a flux $S_{\nu}^{\rm AME} \approx 0.16$ mJy per GBT 0.5' beam if the
source is extended, and potentially larger if the source is smaller
than the 4' CBI beam. For coronene $\Delta
\nu_{\rm line} \approx 340$ MHz and $f_{\rm iso} \approx 6\%$ for the
most abundant isomer, and
using Eq.~(\ref{eq:forecast}) with $\Delta \nu_{\rm tot} = 3.2$ GHz, one may hope to
detect of order $\sim 0.1\%$ of the total PAH abundance concentrated
in the coronene ``family'' with a SNR of $\sim 10$, with $\sim 50$
hours of observations at the GBT. \new{For circumcoronene, with $\Delta
\nu_{\rm line} \approx 70$ MHz and $f_{\rm iso} \approx 1.6\%$, the
same observation would allow to detect approximately 1\% of the total PAH abundance concentrated
in the circumcoronene ``family'' with a SNR of $\sim 10$.}

The \textbf{Perseus molecular cloud}, is a well-known source of AME,
observed near 30 GHz with the VSA \citep{Tibbs_2011} and the \emph{Planck} satellite
\citep{Planck_2011}. The peak emission in the 7' VSA beam is 0.2 Jy,
corresponding to a minimum of 1 mJy per GBT 0.5' beam. If the AME is
truly extended this source is less promising but it is likely that the
emission arises from regions more compact than the 7' VSA beam.

The detection forecasts obtained from Eq.~(\ref{eq:forecast}) may seem
somewhat challenging. However, we emphasise again that
Eq.~(\ref{eq:forecast}) is only an order of magnitude estimate. It is
conceivable that the comb-carrying PAHs are significantly more polar than
the bulk of the AME-emitting PAHs, for example. In addition, when
studying the effect of an asymmetry or inertial defect on the
spectrum, we have assumed conservatively a completely randomised
orientation of the PAH with respect to the angular momentum axis
(i.e. $K$ uniformly distributed between $-J$ and $J$), which is
equivalent to assuming an infinite internal temperature at constant
angular momentum. In practice the internal temperature is finite and
the distribution of $K$-values is somewhat peaked around the lowest
energy states $K = \pm
J$. This reduces the effect of the asymmetry and inertial defect
on the spectrum (in the limit that the grain rotates primarily about its axis of
greatest inertia, its spectrum is a perfect comb no matter how
asymmetric and non-planar it is), and increases the line intensity
with respect to our conservative estimate. Finally, current AME
observation have a larger angular resolution than that of the GBT; we
based our estimate on the conservative assumption that the emission is
extended but the signal may be stronger if the source is in fact compact.

\subsection{A note on protoplanetary discs}

Infrared emission from PAHs is frequently detected in discs around young
stars, in particular Herbig Ae/Be stars, the detection rate being
smaller in the lower mass T-Tauri stars \citep{Acke_2011,
  Kamp_2011}. \cite{Rafikov_2006} has suggested that PAHs may also be
observable through their (continuum) rotational radiation in
circumstellar discs, peaking around 30-50 GHz. Such an observation
would however be made difficult by the presence of other continuum
emission processes, such
as the vibrational emission from large dust grains. It would therefore be very
interesting to search for rotational lines from PAHs in protoplanetary
discs with the method we suggest here. A detection of PAH lines would
allow for identification of the precise species present, and, combined
with spatial information, could shed light on PAH formation and
destruction rates in protoplanetary nebulae. Radio
interferometers such as the VLA\footnote{http://www.vla.nrao.edu/} are
ideally suited for such a search, as they already target
protoplanetary discs at tens of GHz to measure dust properties
\citep{Perez_2012}.

\section{Conclusions} \label{sec:conclusion}

Rotational spectroscopy is a powerful technique to
detect individual molecules in the ISM, but has never been applied to
PAHs with the exception of corannulene. The main deterrent in
searching for the rotational line emission of PAHs has been the
fact that they are a priori large triaxial molecules, for which the
radiated power is diluted among a very large number of weak lines,
making them very difficult to detect. In this paper we have addressed this issue quantitatively, and argued
that the prospects for PAH rotational spectroscopy are brighter than
what may have been foreseen.

It is reasonable to expect that some specific highly-symmetric,
highly-compact PAHs are over-abundant compared to the bulk of the PAH
family. This statement on its own does not solve the problem of
triaxiality: PAHs indeed need a permanent dipole moment in order to
radiate, and symmetric molecules can only be polar if they carry some
impurity, which is bound to break their symmetry. We have therefore
computed the rotational emission of quasi-symmetric, planar
PAHs. Our result is that for a sufficiently small degree of asymmetry,
the rotational spectrum, \emph{observed with a $\sim$ MHz resolution}, has the appearance of a ``comb'' of
evenly spaced strong lines, similar to the spectrum of a much simpler
linear molecule. We have then estimated the degree of asymmetry
resulting from various imperfections that are likely in the ISM. We have concluded that
nitrogen-substituted PAHs are promising targets for rotational
spectroscopy, as they remain very symmetric (if the unsubstituted
species is so), are highly polar, and are moreover
believed to make a significant fraction of the interstellar PAH population.

We have then pointed out that the very peculiar ``comb'' pattern of
the rotational spectrum of quasi-symmetric planar PAHs allows for the use of
matched filtering techniques. This can significantly enhance the
effective signal-to-noise ratio, and allow for detection of PAH
``combs'' hidden in noisy spectra. Moreover, the pattern has a single
free parameter (the line spacing, directly related to the carrier's size) and a blind search over PAHs of
different sizes is therefore possible. This is much more efficient
than targeting a specific species.

Based on the observed level of anomalous microwave emission (AME) in a few
regions, we have made a rough estimate of the fraction of PAHs in
a specific species that could be detected. We have estimated,
assuming conservatively that the emission is extended, that 50 hours of observations of the HII region RCW175
with the GBT could allow for a 10-$\sigma$ detection of about 0.1\% of
the PAH population in coronene and all its substitutes, assuming the observed AME there is due to broadband PAH rotational emission (usually referred to
as spinning dust radiation). 

Since the actual distribution of the PAH
population among specific species (in particular the highly symmetric
ones that can be targeted) is a complete unknown and may
depend on the specific environment, we advocate
for a broad search campaign in various regions known to harbour AME
and/or PAH infrared features. We also point out that protoplanetary disks are
interesting targets for interferometers such as the VLA.

If detected, rotational lines would constitute an unambiguous proof of the existence of free-floating PAHs in the interstellar
medium, conclusively closing the remaining debate about the nature of the carriers
of the aromatic infrared features. The detection of specific PAHs
would also shed light on the formation processes of dust, which the
first step in the path to planet formation.

\section*{Acknowledgements}

I would like to particularly thank Bruce Draine for several enlightening
conversations and detailed comments on the manuscript. I also thank Kfir Blum and Lyman Page for warm
encouragements, Laura P\'erez for discussions on
observational prospects and circumstellar discs, and Chris Hirata for
conversations on quantum chemistry of PAHs. The idea that the discrete rotational line emission from dust grains
may have observational consequences was first discussed during the ``First Billion Years''
workshop organised in August 2010 by the Keck Institute of Space
Studies. This work was supported by the NSF grant number AST-0807444 at the
Institute for Advanced Study.

\appendix

\section{Frequencies of the ``teeth'' of the comb spectrum} \label{appendix}

In this appendix we show that a small asymmetry or inertial defect
change the spacing of the ``comb'' spectrum but do not lead to any
measurable offset, i.e. the frequencies of the ``teeth'' are exactly
$\nu_J = (J + 1/2) \Delta \nu_{\rm line}$, up to corrections much
smaller than a few MHz.

We shall assume that the probability distribution for the total
angular momentum is of the form
\beq
\textrm{Prob}(J) \propto J^2 \exp\left[ - \frac32 \left(\frac{J}{\sigma_J}\right)^2\right],
\eeq
and that $K$ is uniformly distributed in $[-J, J]$, so $\textrm{Prob}(J, K) \propto
\textrm{Prob}(J)/J$ for $J \gg 1$. The power radiated in the $\Delta J
= \Delta K = -1$ transitions at frequency $\nu_{J,K}$ is then, using Eq.~(\ref{eq:J-1Kpm1}).
\beq
\textrm{Pow}(J, K) \propto \nu_{J,K}^4 \left(1 +
  \frac{K}{J}\right)^2 J^{-1}\textrm{Prob}(J).
\eeq
Defining $J_0 \equiv 2 J - K$ so that $\nu_{J, K} \propto J_0$, we
may rewrite this in terms of the variables $J, J_0$ as
\beq
\textrm{Pow}(J, J_0) \propto J_0^4 (3 J - J_0)^2 J^{-1} \exp\left[ - \frac32 \left(\frac{J}{\sigma_J}\right)^2\right].
\eeq
In this expression, $J_0$ labels a line ``stack'' (seen as a single
``tooth'' with finite resolution) whereas $J \in [J_0/3, J_0]$ labels a
specific line within the stack.

The total power per stack is obtained by integrating the above
expressions over $J$. We obtain
\beq
\textrm{Pow}(J_0) \propto J_0^6 \int_{1/3}^1 dj \frac{(3 j - 1)^2}{j}
\exp\left[ -\frac32 \left(\frac{J_0}{\sigma_J}\right)^2 j^2\right].
\eeq
We find numerically that this expression peaks at $J_{0, \rm peak} \approx 2.458~ \sigma_J$.

Let us now consider a fixed stack $J_0$. The density of lines per
frequency interval is
\beq
\rho_{\rm line} (J, J_0) = \left(\frac{d \nu_{J,K}}{d J}\right)^{-1} \propto
\left(1 + \frac38 \frac{\epsilon^2}{\delta} \frac{(J_0-J)^2}{(J_0 - 2 J)^4} J^2 \right)^{-1}.
\eeq

$\bullet$ If $\delta < 0$, this density becomes infinite, i.e. lines accumulate near 
\beq
\frac{J_{\rm acc}}{J_0} = \frac12 \pm \frac1{2 \sqrt{1 + 8 \sqrt{2 \delta / 3 \epsilon^2}}},
\eeq
where there are two physical solutions (corresponding to $J \geq
J_0/3$) if $\delta \geq \frac32 \epsilon^2$ and only one solution
otherwise, the one with the + sign. When there are two solutions one
of them will in general correspond to a larger radiated power
(typically, the + solution, since the spontaneous transition rate of
this transition vanishes at $K  = - J$ and increases towards $K =
+J$). Plugging this value back into Eq.~(\ref{eq:omegaJK}), we see
that lines of the $J_0$-th stack accumulate near the frequency
\beq
\nu_{\rm acc}(J_0) = A_3\left[(2 J_0 + 1) + 2 J_0\left( \epsilon^2
    f(\delta/\epsilon^2) + \delta ~ g(\delta/\epsilon^2)\right)\right], 
\eeq
where $f$ and $g$ are some functions of order unity. We may rewrite
this as
\beq
\nu_{\rm acc}(J_0) = \tilde{A}_3(2 J_0
  + 1) + \mathcal{O}(\epsilon^2, \delta) A_3, \label{eq:nu_acc}
\eeq
where
\beq
\tilde{A}_3 \equiv A_3 \left[1 + \epsilon^2
    f(\delta/\epsilon^2) + \delta ~ g(\delta/\epsilon^2)\right].
\eeq
For $2A_3 \sim 100$ MHz and $\epsilon^2 \sim \delta \sim 10^{-4}$, the
correction term in Eq.~(\ref{eq:nu_acc}) is of order $\sim 10^{-2}$
MHz, much smaller than a resolution bin and is therefore
unobservable. 

$\bullet$ If $\delta > 0$ lines do not accumulate near any particular
frequency, but the radiated power is maximised near some
frequency corresponding to $J_{\rm max}$. In order to find it, we must
maximise $\textrm{Pow}(J, J_0) \rho_{\rm line}(J, J_0)$ at fixed
$J_0$. The result is a function of the form
\beq
J_{\rm max} = J_0 \mathcal{F}(J_0/J_{0, \rm peak}, \epsilon^2/\delta).
\eeq
We solve for $\mathcal{F}$ numerically and find that it is a
relatively slowly varying function of $J_0$ near $J_{0, \rm peak}$: it varies by at most $\sim 10\%$ for $J_0$ within $\pm 15\%$
of $J_{0, \rm peak}$, when $\epsilon^2/\delta$ spans all real
numbers. For reference its value at $J_{0, \rm peak}$ is
$\mathcal{F} \approx 0.5$ for $\epsilon^2 \ll \delta$ to $\mathcal{F}
\approx 1$ (independent of $J_0$) for $\epsilon^2 \gg \delta$.

This implies that the peak frequency within a stack is of the form
\barr
\nu_{\rm peak} &=& A_3\left[(2 J_0 + 1) + 2 J_0\left( \epsilon^2
    f(\delta/\epsilon^2) + \delta ~
    g(\delta/\epsilon^2)\right)\right]\nonumber \\
&+&\mathcal{O}(10\%(\epsilon^2, \delta)) A_3 J_0,
\earr
where the correction is some nonlinear function of $J_0$, which is of
order $\sim 0.1 (\epsilon^2, \delta)$ times the frequency of emission,
that is, for $\nu \approx 30$ GHz and $\epsilon^2, \delta \lesssim
10^{-4}$, the correction is of of order a few tenths of MHz, which is
smaller than the resolution element. Therefore, in this case again, we
conclude that in practice, for narrow enough bandwidth, and for a
resolution of a MHz or more, the comb teeth fall almost exactly at a
constant spacing times $J_0 + 1/2$.

\bibliography{spdust_lines}

\end{document}